\documentclass[prx,twocolumn,preprintnumbers,superscriptaddress]{revtex4-2}
\usepackage{amssymb}
\usepackage{amsmath}
\usepackage{amsfonts}
\usepackage{graphicx}
\usepackage{dcolumn}
\usepackage{bm}
\usepackage{verbatim}
\usepackage{color}
\usepackage{wasysym}
\usepackage{xcolor}
\usepackage{ulem}
\usepackage{appendix}
\usepackage{MnSymbol}

\begin{document}

\title{Two- and many-body physics of ultracold molecules dressed by dual microwave fields}

\author{Fulin Deng}
\thanks{These authors contributed equally to this work.}
\affiliation{CAS Key Laboratory of Theoretical Physics, Institute of Theoretical Physics, Chinese Academy of Sciences, Beijing 100190, China}

\author{Xinyuan Hu}
\thanks{These authors contributed equally to this work.}
\affiliation{CAS Key Laboratory of Theoretical Physics, Institute of Theoretical Physics, Chinese Academy of Sciences, Beijing 100190, China}
\affiliation{School of Physical Sciences, University of Chinese Academy of Sciences, Beijing 100049, China}

\author{Wei-Jian Jin}
\thanks{These authors contributed equally to this work.}
\affiliation{CAS Key Laboratory of Theoretical Physics, Institute of Theoretical Physics, Chinese Academy of Sciences, Beijing 100190, China}
\affiliation{School of Physical Sciences, University of Chinese Academy of Sciences, Beijing 100049, China}

\author{Su Yi}
\email{yisu@nbu.edu.cn}
\affiliation{Institute of Fundamental Physics and Quantum Technology $\&$ School of Physics, Ningbo University, Ningbo, 315211, China}
\affiliation{Peng Huanwu Collaborative Center for Research and Education, Beihang University, Beijing 100191, China}

\author{Tao Shi}
\email{tshi@itp.ac.cn}
\affiliation{CAS Key Laboratory of Theoretical Physics, Institute of Theoretical Physics, Chinese Academy of Sciences, Beijing 100190, China}
\affiliation{School of Physical Sciences, University of Chinese Academy of Sciences, Beijing 100049, China}

\date{\today }

\begin{abstract}
We investigate the two- and many-body physics of the ultracold polar molecules dressed by dual microwaves with distinct polarizations. Using Floquet theory and multichannel scattering calculations, we identify a regime with the largest elastic-to-inelastic scattering ratio which is favorable for performing evaporative cooling. Furthermore, we derive and, subsequently, validate an effective interaction potential that accurately captures the dynamics of microwave-shielded polar molecules (MSPMs). We also explore the ground-state properties of the ultracold gases of MSPMs by computing physical quantities such as gas density, condensate fraction, momentum distribution, and second-order correlation. It is shown that the system supports a weakly correlated expanding gas state and a strongly correlated self-bound gas state. Since the dual-microwave scheme introduces addition control knob and is essential for creating ultracold Bose gases of polar molecules, our work pave the way for studying two- and many-body physics of the ultracold polar molecules dressed by dual microwaves.
\end{abstract}

\maketitle

\section{Introduction}

The realization of ultracold molecular gases has become a cornerstone for harnessing the potential of quantum technologies, including advancements in precision measurements~\cite{Kozlov2007,Berger2010,Hinds2011,ThO2014,ThO2018,Hutzler2020}, ultracold chemistry~\cite{Krem2008,Ni2019,Liu2022}, quantum computing~\cite{DeMille2002,Cornish2020,Zoller2006a,Tesch2002,Wall2015,Albert2020}, and quantum simulation~\cite{Zoller2006,Zwierlein2021,Pfau2009}. Recent progresses in microwave shielding techniques~\cite{Karman2018,Quemener2018,Ye2019,Luo2021,Doyle2021,Wang2023,Will2023} have enabled the creation of stable degenerate Fermi gases of NaK molecules~\cite{Luo2022a,Luo2022b} and Bose-Einstein condensates (BECs) of NaCs molecules~\cite{Will2023b}. These breakthroughs highlight the transformative potential of microwave-shielded polar molecules (MSPMs) in both fundamental and applied quantum science.

For fermionic molecules like NaK, a circularly polarized ($\sigma^+$) microwave field generates tunable dipole-dipole interactions (DDIs) and creates a robust shielding potential that suppresses losses via the formation of four-body complexes. With Pauli blocking effectively reducing three-body losses, this shielding facilitates evaporative cooling to achieve ultracold NaK tetramers~\cite{chen2023,Deng2024}. This opens avenues for realizing $p$-wave superfluidity~\cite{You1999,Baranov2002,Shi2010,Pu2010,Hirsch2010, Shlyapnikov2011,Baranov2012,Shi2014,Zhai2013,Deng2023}, a critical component for topological quantum computing~\cite{Read2000,Ivanov2001}. Conversely, bosonic molecules encounter significant challenges due to strong attractive dipolar interactions under a $\sigma^+$-polarized microwave field, which intensify three-body losses and impede BEC formation~\cite{Wang2023}. This obstacle has been successfully overcome experimentally using a dual-microwave approach, where an additional linearly polarized ($\pi$) microwave field reduces the attractive forces, substantially suppressing three-body losses~\cite{Will2024} and enabling the formation of stable NaCs molecular BECs~\cite{Will2023b}. In addition, the ground-state properties of the ultracold Bose gases of MSPMs are theoretically investigated~\cite{Jin2024,Langen2024}.

Despite these experimental success, the underlying shielding mechanisms remains incompletely understood since, in the presence of the second microwave, the time dependence of DDIs cannot be eliminated through rotations, which poses a serious challenge to theoretical treatments. Furthermore, dual microwave fields enable molecules to absorb photons from one field and emit photons to the other, resulting in an energy exchange process that drives inelastic scattering and heating. Thus the dual-microwave scheme may jeopardize the evaporative cooling. To achieve an optimal balance between suppressing inelastic losses while preserving elastic scattering and shielding, it is necessary to carry out a detailed Floquet theoretical analysis of scattering dynamics governed by a time-dependent Hamiltonian. 

In this study, we develop a universal theoretical framework to explore the scattering and many-body physics of polar molecules under dual microwave fields. By integrating Floquet theory with multichannel scattering calculations for time-dependent interactions, we compute (in)elastic scattering rates and identify regimes where the elastic-to-inelastic scattering ratio is maximized. In this regime, the evaporative cooling efficiency can be significantly enhanced, in agreement with the experimental choice. Moreover, to elucidate the shielding mechanism and characterize the key interaction dynamics of MSPMs, we analytically derive the effective potential between MSPMs by taking into account all second-order contributions in the Floquet formulation. The validity of the effective potential is thoroughly checked via the multi-channel scattering calculations. Finally, we explore the ground-state properties of an ultracold NaCs gas by employing a variational wave function incorporating with Jastrow correlation~\cite{Jin2024}. By tuning the Rabi frequency of the $\pi$-polarized microwave, we find a weakly correlated expanding gas state and a strongly correlated self-bound gas state, in analogy to the states found in the ultracold NaRb gases~\cite{Jin2024,Langen2024}. The condensate fractions and momentum distributions of both states are also computed. 

This work is organized as follows. In Sec.~\ref{secmod}, we present our model for two interacting molecules subjected to a dual microwaves. In Sec.~\ref{secfloq}, we first formulate the two-body problem in the framework of the Floquet theory which can be used for the multi-channel calculations. We then derive an effective potential for MSPMs and propose a single-channel model. In Sec.~\ref{secresu}, we present our results on the two- and many-body physics. Finally, we conclude in Sec.~\ref{secconcl}

\section{Model}\label{secmod}

\begin{figure}[tbp]
\includegraphics[width=\linewidth]{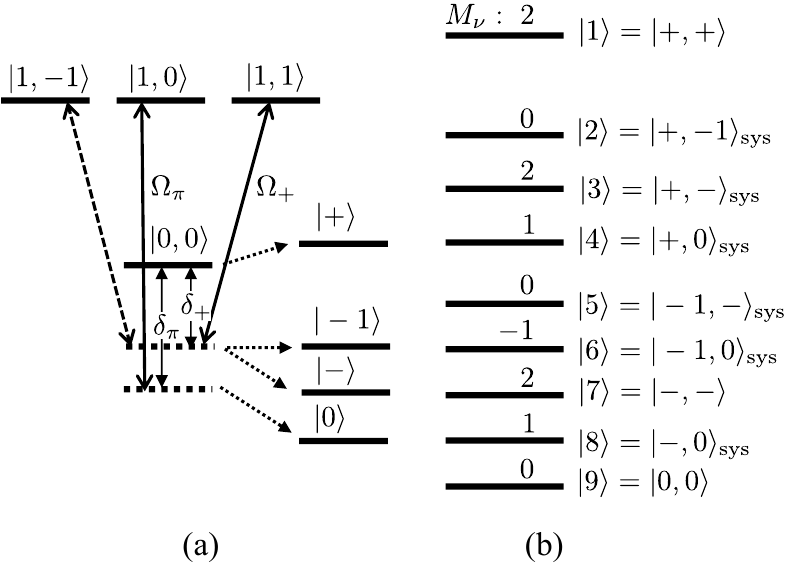}
\caption{Schematics of the dressed-state energy levels. (a) Single-molecule states. (b) Two-molecule states in the symmetric subspace. $M_\nu$ is the quantum number associated with the two-molecules states [see Eq.~\eqref{SR}].}
\label{levels}
\end{figure}

We consider an ultracold gas of the bialkali polar molecules in the $^1\Sigma(v=0)$ state which can be treated as a rigid rotor. Under ultracold temperature, we may focus on the lowest ($|J,M_{J}\rangle=|0,0\rangle $) and the first excited ($|J=1,M_{J}=1,0,-1\rangle $) rotational manifolds which are split by the energy $\hbar\omega_{e}$. Each molecule possesses an electric dipole moment $d\hat{\boldsymbol d}$ which couples to the external fields. Here $d$ is the permanent dipole moment in the molecular frame and $\hat{\boldsymbol d}$ is the unit vector along the internuclear axis of the molecule. To achieve shielding with dual microwaves, a $\sigma^{+}$- and a $\pi$-polarized microwaves are applied to couple the transitions $|0,0\rangle\leftrightarrow |1,1\rangle $ and $|0,0\rangle \leftrightarrow |1,0\rangle $, respectively. The frequencies of the $\sigma ^{+}$ and $\pi$ microwaves, i.e., $\omega_{+}$ and $\omega_{\pi}$, respectively, are blue detuned from the transition frequency $\omega_e$. In the frame co-rotating with microwave fields, the internal-state Hamiltonian for a single molecule is time-independent and takes the form
\begin{align}
\hat{h}_{\mathrm{in}}&=\delta_+\Big(|1,1\rangle\langle1,1|+|1,-1\rangle%
\langle 1,-1|\Big)+\delta_\pi|1,0\rangle\langle 1,0|  \notag \\
&\quad+\left(\frac{\Omega_+}{2}|1,1\rangle\langle 0,0|+\frac{\Omega_\pi}{2}%
|1,0\rangle\langle 0,0| +\mathrm{H.c.}\right),
\end{align}%
where $\Omega_{+}$ ($\Omega_{\pi}$) and $\delta_{+}=\omega_{e}-\omega_{+}$ ($%
\delta_{\pi }=\omega_{e}-\omega_{\pi}$) are the Rabi frequency and detuning
of the $\sigma^+$ ($\pi$) microwave, respectively.

The single-molecule Hamiltonian $\hat{h}_{\mathrm{in}}$ can be analytically diagonalized by an unitary transformation $U_{1}(\alpha,\beta,\gamma)$ which is conveniently parameterized by three Euler angles $\alpha $, $\beta $, and $\gamma $ (see App.~\ref{appsingle} for details). The columns of $U_{1}$ are the eigenvectors of $\hat{h}_{\mathrm{in}}$ which, from left to right, are denoted as $\left\vert +\right\rangle$, $\left\vert -1\right\rangle(\equiv\left\vert 1,-1\right\rangle)$, $\left\vert -\right\rangle$, and $\left\vert 0\right\rangle$ [see Fig.~\ref{levels}(a)]. The corresponding eigenenergies are denoted as $E_{+}$, $E_{-1}=\delta_{+}$, $E_{-}$, and $E_{0}$, respectively. We note that in the limit $\Omega _{\pi }\rightarrow 0$ the eigenstates and eigenenergies can be expressed explicitly as $|+ \rangle\rightarrow \cos \alpha \left\vert 0,0\right\rangle + \sin \alpha\left\vert 1,1\right\rangle $, $|- \rangle\rightarrow \sin \alpha \left\vert 0,0\right\rangle - \cos \alpha\left\vert 1,1\right\rangle $, $|0\rangle\rightarrow |1,0\rangle$, $E_{\pm}\rightarrow (\delta _{+}\pm \Omega _{\mathrm{eff}})/2$, and $E_{0}\rightarrow \delta _{\pi }$, where $\Omega _{\mathrm{eff}}=\sqrt{\delta _{+}^{2}+\Omega_{+}^{2}}$ and the Euler angles are known analytically, i.e., $(\alpha ,\beta ,\gamma )=\left(\arccos [(1-\delta _{+}/\Omega _{\mathrm{eff}})/2]^{1/2},0,0\right)$.

Now, for two molecules with dipole moments $d\hat{\boldsymbol d}_{1}$ and $d\hat{\boldsymbol d}_{2}$, the inter-molecular DDI is
\begin{align}
V_{\mathrm{dd}}({\boldsymbol r})& =\frac{d^{2}}{4\pi \epsilon _{0}r^{3}}\left[
\hat{\boldsymbol d}_{1}\cdot \hat{\boldsymbol d}_{2}-3(\hat{\boldsymbol d}_{1}\cdot
\hat{\boldsymbol r})(\hat{\boldsymbol d}_{2}\cdot \hat{\boldsymbol r})\right]  \notag\\
& =-\frac{\eta}{r^3}
\sum_{m=-2}^{2}Y_{2m}^{\ast }(\hat{\boldsymbol r})\Sigma _{2,m}
\end{align}%
where $\eta=\sqrt{8\pi/15}\,d^2/\epsilon_{0}$ with $\epsilon_{0}$ being the electric permittivity of vacuum, $r=|{{\boldsymbol r}%
}|$, $Y_{2m}(\hat{\boldsymbol r})$ are spherical harmonics, and $\Sigma _{2,m}$ are components of the rank-2 spherical tensor defined as $\Sigma _{2,0}=(\hat{%
d}_{1}^{+}\hat{d}_{2}^{-}+\hat{d}_{1}^{-}\hat{d}_{2}^{+}+2\hat{d}_{1}^{0}%
\hat{d}_{2}^{0})/\sqrt{6}$, $\Sigma _{2,\pm 1}=(\hat{d}_{1}^{\pm }\hat{d}%
_{2}^{0}+\hat{d}_{1}^{0}\hat{d}_{2}^{\pm })/\sqrt{2}$, and $\Sigma _{2,\pm
2}=\hat{d}_{1}^{\pm }\hat{d}_{2}^{\pm }$ with $\hat{d}_{j}^{\pm }=Y_{1,\pm
1}(\hat{\boldsymbol d}_{j})$ and $\hat{d}_{j}^{0}=Y_{1,0}(\hat{\boldsymbol d}_{j})$. In the rotating frame, $\hat{d}_{j}^{\pm }$ and $\hat{d}_{j}^{0}$ become time-dependent. In fact, in the basis $|J,M_{J}\rangle $, these
operators can be written out explicitly as
\begin{align}
\hat{d}_{j}^{0}& =\frac{1}{\sqrt{4\pi }}\left( |0,0\rangle \langle
1,0|e^{-i\omega _{\pi }t}+\mathrm{h.c.}\right) ,  \notag \\
\hat{d}_{j}^{+}& =\frac{1}{\sqrt{4\pi }}\left( -|0,0\rangle \langle
1,-1|e^{-i\omega _{+}t}+|1,1\rangle \langle 0,0|e^{i\omega _{+}t}\right) ,
\notag \\
\hat{d}_{j}^{-}& =-\left( \hat{d}_{j}^{+}\right) ^{\dag }.  \notag
\end{align}%
After substituting the above spherical components of the vector into $\Sigma
_{2m}$, we find
\begin{widetext}
\begin{align}
\Sigma _{2,0} &=\frac{1}{4\pi \sqrt{6}}\Big(2\left\vert 1,0\right\rangle
\left\langle 0,0\right\vert \otimes \left\vert 0,0\right\rangle \left\langle
1,0\right\vert -\left\vert 1,1\right\rangle \left\langle 0,0\right\vert
\otimes \left\vert 0,0\right\rangle \left\langle 1,1\right\vert -\left\vert
0,0\right\rangle \left\langle 1,-1\right\vert \otimes \left\vert
1,-1\right\rangle \left\langle 0,0\right\vert +\mathrm{h.c.}\Big),  \notag \\
\Sigma _{2,1}&=\frac{1}{4\pi \sqrt{2}}\Big[\big(\left\vert 1,1\right\rangle
\left\langle 0,0\right\vert \otimes \left\vert 0,0\right\rangle \left\langle
1,0\right\vert +\left\vert 0,0\right\rangle \left\langle 1,0\right\vert
\otimes \left\vert 1,1\right\rangle \left\langle 0,0\right\vert \big)e^{i\omega t}\nonumber\\
&\qquad\qquad\;-\big(\left\vert 0,0\right\rangle
\left\langle 1,-1\right\vert \otimes \left\vert 1,0\right\rangle
\left\langle 0,0\right\vert +\left\vert 1,0\right\rangle \left\langle
0,0\right\vert \otimes \left\vert 0,0\right\rangle \left\langle
1,-1\right\vert \big)e^{-i\omega t}\Big],  \notag \\
\Sigma _{2,2} &=-\frac{1}{4\pi }\big(\left\vert 1,1\right\rangle \left\langle
0,0\right\vert \otimes \left\vert 0,0\right\rangle \left\langle
1,-1\right\vert +\left\vert 0,0\right\rangle \left\langle 1,-1\right\vert
\otimes \left\vert 1,1\right\rangle \left\langle 0,0\right\vert \big),\nonumber
\end{align}
\end{widetext}
where, according to the rotating-wave approximation, we have neglected time-dependent terms with higher frequencies $\omega _{+}$ and $\omega _{\pi }$ (of the order of $\mathrm{GHz}$) and retained those with the lower frequency $\omega =\omega _{+}-\omega_{\pi }$.

To proceed further, the Hamiltonian for the relative motion of two molecules is
\begin{equation*}
\hat{H}_{2}=-\frac{\hbar ^{2}\nabla ^{2}}{M}+\sum_{j=1,2}\hat{h}_{\mathrm{in}%
}(j)+V_{\mathrm{dd}}({\boldsymbol r},t),
\end{equation*}%
where $M$ is the mass of the molecule and $\hat{h}_{\mathrm{in}}(j)$ denotes
the internal-state Hamiltonian of the $j$th molecule. And we have explicitly expressed $V_{\mathrm{dd}}$ as a function of $t$ in the rotating frame. Since $\hat{H}_{2}$ possesses a parity symmetry, the
symmetric and antisymmetric two-particle internal states are decoupled in
the Hamiltonian $\hat{H}_{2}$. We shall only focus on the ten-dimensional
symmetric subspace in which the microwave shielded two-molecule state $%
|1\rangle \equiv |+\rangle\otimes |+\rangle$ lies. Further simplification can be made by noting that $|1\rangle $ only couples to the following eight two-molecule states [see the schematic plot in Fig.~\ref{levels}(b)]: $|2\rangle \equiv|+,-1\rangle _{\mathrm{sys}}$, $|3\rangle \equiv |+,-\rangle_{\mathrm{sys}}$, $|4\rangle \equiv |+,0\rangle _{\mathrm{sys}}$, $|5\rangle \equiv |-1,-\rangle _{\mathrm{sys}}$, $|6\rangle \equiv|-1,0\rangle_{\mathrm{sys}}$, $|7\rangle \equiv |-\rangle\otimes|-\rangle$, $|8\rangle =|-,0\rangle_{\mathrm{sys}}$, and $|9\rangle \equiv|0\rangle\otimes |0\rangle$, where $|i,j\rangle_{\mathrm{sys}}=(|i\rangle\otimes |j\rangle+|j\rangle\otimes |i\rangle)/\sqrt{2}$ represents the symmetrized two-molecule state. The corresponding energies of these two-particle states are denoted as $E_\nu^{(\infty)}$. As a result,
these nine two-molecule states form a 9-dimensional (9D) symmetric subspace,
$\mathcal{S}_{9}\equiv \mathrm{span}\{|\nu \rangle \}_{\nu =1}^{9}$. Now,
since we focus on system with all molecules being prepared in the microwave shielded $|+\rangle$ state, we may project the interaction $V_{\mathrm{dd}}({\boldsymbol r})$ onto the two-molecule subspace $\mathcal{S}_{9}$.

In an attempt to eliminate the time dependence of the $\hat{H}_{2}$, we introduce, in $\mathcal{S}_{9}$, an unitary transformation defined by the diagonal matrix,
\begin{equation}
U_{2}(t)=\mathrm{diag}\left( 1,1,1,e^{-i\omega t},1,e^{-i\omega
t},1,e^{-i\omega t},e^{-2i\omega t}\right) .
\end{equation}%
A straightforward calculation shows that the Hamiltonian $\hat{H%
}_{2}$ is transformed into
\begin{eqnarray}
\hat{\mathcal{H}}(t) &=&U_{2}^{\dag }(t)\hat{H}_{2}U_{2}(t)-iU_{2}^{\dagger
}(t)\partial _{t}U_{2}(t)  \notag \\
&=&-\frac{\hbar ^{2}\nabla ^{2}}{M}+\mathcal{E}^{(\infty )}+\mathcal{V}({%
{\boldsymbol r}},t),
\end{eqnarray}%
where $\mathcal{E}^{(\infty)}$ is a diagonal matrix with elements being the energies of the asymptotical state $|\nu\rangle$ with respect to that of the $|\nu=1\rangle$ state, i.e., $\mathcal{E}^{(\infty)}_{\nu\nu'}=(E_\nu^{(\infty)}-2E_+)\delta_{\nu\nu'}$. Moreover, the two-body interaction in the basis $\{|\nu\rangle\}$ is
\begin{equation*}
\mathcal{V}({\boldsymbol r},t)=U_{2}^{\dag }(t)V_{\mathrm{dd}}({\boldsymbol r}%
)U_{2}(t)=\sum_{s=-3}^{3}\mathcal{V}_{s}({\boldsymbol r})e^{is\omega t}
\end{equation*}%
which is decomposed into components according to the time dependence $e^{is\omega t}$. Particularly, the components satisfy $\mathcal{V}_{-s}({\boldsymbol r})=\mathcal{V}_{s}^{\dagger }({\boldsymbol r})$ and 
\begin{subequations}\label{2bodyint}
\begin{align}
\mathcal{V}_{0}({\boldsymbol r}) &=-\frac{\eta}{r^{3}}\left[Y_{20}(\hat{\boldsymbol r})\Sigma
_{2,0}^{(0)}+Y_{21}^{\ast }(\hat{\boldsymbol r})\Sigma _{2,1}^{(0)}+Y_{21}(\hat{\boldsymbol r})\Sigma _{2,1}^{(0)\dagger }\right.\nonumber\\
&\qquad\quad\;\;\left.+Y_{22}^{\ast }(\hat{\boldsymbol r})\Sigma _{2,2}^{(0)}+Y_{22}(\hat{\boldsymbol r})\Sigma _{2,2}^{(0)\dagger }\right],\\
\mathcal{V}_{1}({\boldsymbol r}) &=-\frac{\eta}{r^{3}}\left[Y_{20}(\hat{\boldsymbol r})\Sigma _{2,0}^{(1)}+Y_{21}(\hat{\boldsymbol r})\Sigma_{2,1}^{(-1)\dag}+Y_{21}^*(\hat{\boldsymbol r})\Sigma _{2,1}^{(1) }\right.\nonumber\\
&\qquad\quad\;\;\left.+Y_{22}(\hat{\boldsymbol r})\Sigma_{2,2}^{(-1)\dag}+Y_{22}^*(\hat{\boldsymbol r})\Sigma _{2,2}^{(1)}\right], \\
\mathcal{V}_{2}({\boldsymbol r}) &=-\frac{\eta}{r^{3}}\left[Y_{20}(\hat{\boldsymbol r})\Sigma _{2,0}^{(2)}+Y_{21}(\hat{\boldsymbol r})\Sigma _{2,1}^{(-2)\dag}\right.\nonumber\\
&\qquad\quad\;\;\left.+Y_{21}^*(\hat{\boldsymbol r})\Sigma _{2,1}^{(2) }+Y_{22}^*(\hat{\boldsymbol r})\Sigma_{2,2}^{(2)}\right],   \\
\mathcal{V}_{3}({\boldsymbol r}) &=-\frac{\eta}{r^{3}}Y_{21}^*(\hat{\boldsymbol r})\Sigma _{2,1}^{(3)},
\end{align}
\end{subequations}
where $\Sigma_{2,m}^{(s)}$ are matrices originated from $\Sigma_{2,m}$ that is associated with the time dependence $e^{is\omega t}$. For completeness, the matrix elements of $\Sigma_{2,m}$ are listed in the App.~\ref{appmatele}, which satisfy $\Sigma _{2,-m}^{(s)}=(-1)^{m}\Sigma_{2,m}^{(-s)\dag }$.

Although the expression for the interaction Hamiltonian $\mathcal{V}$ may appear very complicated, each term has a clear physical interpretation. To see the physical processes associated with the interaction, we explicitly write out the interaction Hamiltonian in the basis $\{|\nu \rangle\}$, namely,
\begin{equation}
-\frac{\eta }{r^{3}}\sum_{sm}\sum_{\nu \nu
^{\prime }}\left( \Sigma _{2,m}^{(s)}\right) _{\nu \nu ^{\prime
}}Y_{2m}^{\ast }(\hat{\boldsymbol r})e^{is\omega t}|\nu \rangle \langle \nu ^{\prime }|.
\label{Vc}
\end{equation}
Next, we assign each single-molecule state a quantum number, i.e., $\{1,-1,1,0\}$ to $\{|+\rangle, |-1\rangle, |-\rangle, |0\rangle\}$, respectively, which represents the $z$ component of the angular momentum. Then the quantum numbers associated with the two-molecule states $|\nu\rangle$ are $M_{\nu}=2,0,2,1,0,-1,2,1$, and $0$ for $\nu=1$ to $9$, respectively [see Fig.~\ref{levels}(b)]. Now, by inspection, we find matrix element $\left( \Sigma_{2,m}^{(s)}\right) _{\nu \nu ^{\prime }}$ is nonzero only when the selection rule
\begin{equation}
M_{\nu }=M_{\nu ^{\prime }}+m-s  \label{SR}
\end{equation}%
is satisfied. To proceed further, we interpret the phase factor $e^{is\omega t}=e^{is\omega
_{+}t}e^{-is\omega _{\pi }t}$ in Eq.~\eqref{Vc} as the creation of $s$ $\sigma^+$-microwave photons and the annihilation of $s$ $\pi$-microwave photons. As the result, the net change of the quanta associated with the polarization of the microwave photons is $s$. It is now clear that the selection rule~\eqref{SR} explicitly represents a change in internal angular momenta (i.e., $M_{\nu^{\prime}} \rightarrow M_\nu$) during the transition from the state $|\nu ^{\prime }\rangle $ to the state $|\nu \rangle$ via the emission of $s$ circular polarizations and the absorption of $m$ angular momentum quanta from the orbital motion. Thus, the selection rule (\ref{SR}) reveals a conservation of total angular momentum, i.e., the sum of microwave polarizations, the angular momentum of internal states $|\nu \rangle$, and orbital angular momentum, projected along the $z$ axis.

\section{Multi-channel scatterings for Floquet-Bloch states}\label{secfloq}

The time periodicity of the Hamiltonian $\hat{\mathcal{H}}(t)$ suggests that we may tackle the two-molecule physics using the Floquet theory. Specifically, the solution of the Schr\"{o}dinger equation,
\begin{equation}
i\hbar \frac{\partial |\psi (t)\rangle }{\partial t}=\hat{\mathcal{H}}%
(t)\left\vert \psi (t)\right\rangle ,  \label{se}
\end{equation}%
takes the \textquotedblleft Floquet-Fourier\textquotedblright\ form
\begin{equation}
\left\vert \psi (t)\right\rangle =e^{-i\varepsilon t}\sum_{n=-\infty
}^{\infty }e^{-in\omega t}\left\vert \psi _{n}\right\rangle ,
\end{equation}%
where $\varepsilon $ is the quasi-energy of the state and $|\psi _{n}\rangle $ is the time-independent harmonic component defined on the 9D Hilbert space $\mathcal{S}_9$. It follows from Eq.~(\ref{se}) that $|\psi _{n}\rangle $ satisfy the time-independent eigenvalue equation:
\begin{align}
\sum_{s}\mathcal{H}_{s}\left\vert \psi _{n+s}\right\rangle-n\omega|\psi_n\rangle=\varepsilon \left\vert \psi _{n}\right\rangle ,\quad n=0,\pm 1,\pm 2,\ldots
\label{FE}
\end{align}%
where the Floquet Hamiltonian is
\begin{equation*}
\mathcal{H}_{s}=\left\{
\begin{array}{ll}
-\hbar^2\nabla^2/M+\mathcal{E}^{(\infty )}+\mathcal{V}_{0}({\boldsymbol r}), & \mbox{for }s=0; \\
\mathcal{V}_{s}({\boldsymbol r}), & \mbox{for }0<|s|\leq 3; \\
0, & \mbox{otherwise}.
\end{array}%
\right.
\end{equation*}
Introducing vector $|{\boldsymbol\Psi}\rangle=(\cdots,|\psi_{-1}\rangle,|\psi_{0}\rangle,|\psi_{1}\rangle,\cdots)^T$ for the Floquet space wavefuction, Eq.~\eqref{FE} can be rewritten into a more compact form as
\begin{align}
{\mathbf H}|{\boldsymbol\Psi}\rangle=\varepsilon|{\boldsymbol\Psi}\rangle,\label{FE2}
\end{align}
where, in terms of $9\times9$ block matrices, the Floquet Hamiltonian ${\mathbf H}$ is a heptadiagonal matrix. 
Thus one may easily visualize the structure of the Schr\"odinger Eq.~\eqref{FE} through $\mathbf H$.

With Eq.~\eqref{FE}, the advantage of performing the transformation $U_{2}(t)$ is now understandable. Specifically, in the absence of the $\pi $-field, $U_{2}(t)$ leads to a time-independent $\hat{\mathcal{H}}$, i.e., $\mathcal{H}_{s}=0$ for all $s\neq 0$, indicating that different Floquet sectors are decoupled and the eigenstates can be obtain directly by diagonalizing $\mathcal{H}_{0}$. Then as the $\pi$-polarized microwave is gradually turned on to lower the attractive interaction on the $xy$ plane, the transitions between different Floquet sectors are also switched on. Moreover, to maintain the shielding effect along all directions, the Rabi frequency of the $\pi$-polarized microwave has to be much smaller than that of the $\sigma^+$-polarized microwave (see Sec.~\ref{EP} for details). As a result, through transformation $U_2$, we ensure that the transitions between different Floquet sectors in the presence of the $\pi$ microwave are perturbation. Such structure accelerates the convergence in numerical calculations.

Now, we turn to formulate the multi-channel scatterings for Floquet-Bloch states. To solve Eq.~\eqref{FE}, we expand the eigenstate wavefunction $\psi_{n,\nu }({\boldsymbol r})=\langle \nu|\psi_{n}({\boldsymbol r})\rangle$ in the partial-wave basis as
\begin{align}
\psi_{n,\nu }({\boldsymbol r})=\sum_{lm_\nu}r^{-1}\phi_{n\nu l}(r)Y_{l,m_{\nu }+n}(\hat{\boldsymbol r}),\label{expansion}
\end{align}
where $\phi_{n\nu l}$ are the radial wavefunctions. Interestingly, due to the total angular momentum conservation, $m_\nu$ for different interaction channels are fixed as follows. Given $m_1=m_0$, then, for $\nu=2$ to $9$, we have $m_\nu=m_0+2,m_0,m_0+1,m_0+2,m_0+3,m_0,m_0+1$, and $m_0+2$, respectively. More importantly, since different sets of $\{m_\nu\}_{\nu=1}^9$ are decoupled, we may drop the summation over $m_\nu$ in Eq.~\eqref{expansion} and consider each set separately. This observation greatly simplifies the numerical calculations. 

To proceed further, it is convenient to introduce the column vector ${\boldsymbol\Phi}({\boldsymbol r})$ formed by the elements $\phi_{n\nu l}(r)$. The Schr\"odinger equation for the radial wavefunction can then be written as
\begin{align}
\partial_{r}^{2}{\boldsymbol\Phi} (r)+\mathbf{W}(r){\boldsymbol\Phi} (r)=0,  \label{SE2}
\end{align}
where the matrix $\mathbf{W}$ is defined by the elements
\begin{align}
W_{n\nu l,n'\nu' l'}(r) &=\left[k_{n\nu}^{2}-\frac{l(l+1)}{r^{2}}\right]\delta_{nn^{\prime }}\delta _{\nu \nu ^{\prime}}\delta _{ll^{\prime }}\nonumber\\
&\quad-M\sum_{s=-3}^{3}\delta_{n+s,n^{\prime }}V_{\nu l,\nu ^{\prime }l^{\prime
}}^{(n,s)}(r)
\end{align}
with $k_{n\nu}=\left[M\left(\varepsilon-\mathcal{E}_{\nu\nu}^{(\infty)}+n\omega\right)\right]^{1/2}$ being the incident momentum with respect to the $n$th Floquet sector and the $\nu$th interaction channel and
\begin{align}
V_{\nu l,\nu ^{\prime }l^{\prime }}^{(n,s)}(r)=\int d\hat{\boldsymbol r} Y_{l,m_{\nu
}+n}^{\ast }(\hat{{\boldsymbol r}})\left[\mathcal{V}_{s}({\boldsymbol r})\right]_{\nu\nu'}Y_{l^{\prime },m_{\nu
^{\prime }}+n+s}(\hat{{\boldsymbol r}})
\end{align}%
which can be evaluated analytically.

To solve the multi-channel scattering problem, we numerically evolve Eq.~\eqref{SE2} from a ultraviolet cutoff $r_{\rm{UV}}$ to a sufficiently large value $r_\infty$ using Johnson’s log-derivative propagator method. Here, we impose a capture boundary condition~\cite{Karman2018} at $r_{\rm{UV}}$. It turns out that the choice of $r_{\rm{UV}}$ does not affect the result if it is deep inside the shielding core~\cite{Deng2023,Deng2024}.
We remark that to account for the short-range effects in scatterings, we also include, in numerical calculations, the universal van der Waals interaction through the replacement $\mathcal{V}_{0}({\boldsymbol r})\rightarrow\mathcal{V}_{0}({\boldsymbol r})-C_{\mathrm{vdW}}/r^{6}$, where $C_{\mathrm{vdW}}$ is the strength of the universal van der Waals interaction~\cite{Julienne2010}. Since $C_{\mathrm{vdW}}$ is generally much smaller than the microwave shielding strength $C_6$~\cite{Lepers2013}, the $C_{\mathrm{vdW}}$ term only takes effect at short distance. Then to obtain the scattering matrix, we compare $\phi_{n\nu l}(r_\infty)$ with the asymptotic boundary condition
\begin{align}
\phi_{n\nu l}^{(m_0)}(r)&\stackrel{r\rightarrow\infty}{\longrightarrow}(k_{n_{0}\nu _{0}}r)^{-1/2}\hat{j}_{l}(k_{n_{0}\nu _{0}}r)\delta_{nn_{0}}\delta _{\nu \nu _{0}}\delta _{ll_{0}}\nonumber\\
&\quad\;\;\;\;\;+(k_{n\nu}r)^{-1/2}\hat{n}_{l}(k_{n\nu}r)K_{n\nu l,n_{0}\nu _{0}l_{0}}^{(m_0)},
\end{align}
where $\hat j_l(z)$ and $\hat n_l(z)$ are the Riccati-Bessel functions, and $K_{n\nu l,n_{0}\nu _{0}l_{0}}^{(m_0)}$ are elements of the $K$ matrix, corresponding to the scattering from the incident channel $(n_0\nu_0l_0)$ to the outgoing one $(n\nu l)$. Here both channels are characterized by the same projection quantum number $m_0$. Moreover, $k_{n_{0}\nu _{0}}$ and $k_{n\nu}$ are the relative momenta for the incident and outgoing channels, respectively. In numerical calculations, we introduce a truncation $n_{\rm cut}$ for the Floquet Hamiltonian such that $|n|\leq n_{\rm cut}$. Practically, it is found that, for control parameters covered in this work, the scattering solutions converge when $n_{\rm cut}=5$.

To proceed further, we denote the $K$ matrix as ${\mathbf K}^{(m_0)}$, from which one can obtain the scattering matrix $\mathbf{S}^{(m_0)}=\left(1+i\mathbf{K}^{(m_0)}\right)\left(1-i\mathbf{K}^{(m_0)}\right)^{-1}$. 
Now, the total elastic cross section for the incident channel $(n_0\nu_0)$ is
\begin{align}
\sigma_{n_0\nu_0}^{(\mathrm{el})} &=\frac{2\pi }{k_{n_0\nu_0}^{2}}\sum_{ll_{0}m_0}\left|\delta_{ll_{0}} -S_{n_{0}\nu_{0}l,n_{0}\nu_{0}l_{0}}^{(m_0)}\right|^{2},
\end{align}
where $S_{n\nu l,n_{0}\nu_{0}l_{0}}^{(m_0)}$ are the elements of the scattering matrix and $l$ and $l_0$ are even (odd) for bosons (fermions). For elastic scattering, the outgoing channel of the molecules is the same as incident channel $(n_0\nu_0)$ and the total kinetic energy of the molecules is thus conserved. Next, the total inelastic cross section can be calculated by subtracting the total elastic cross section from the total cross section, i.e.,
\begin{align}
\sigma_{n_0\nu_0}^{(\mathrm{inel})} &=\frac{2\pi}{k_{n_0\nu_0}^{2}}\sum_{ll_{0}m_0}\left(\delta
_{ll_{0}}-\left|S_{n_{0}\nu _{0}l,n_{0}\nu _{0}l_{0}}^{(m_0)}\right|^{2}\right).\label{eqinel}
\end{align}
It is instructive to distinguish, depending on whether the total energy of the colliding molecules is conserved, the degenerate and nondegenerate inelastic scatterings which are disguised in  Eq.~\eqref{eqinel}. Specifically, for degenerate inelastic scatterings, the outgoing molecules remain in the same Floquet sector $n_0$ but transit to the lower dressed-state channel $\nu$ ($>\nu_0$). Thus the total energy of the colliding molecules is conserved. While for nondegenerate inelastic scatterings, the outgoing molecules transit to a distinct Floquet sector $n$ ($\neq n_0$) by absorbing or emitting microwave photons and thus the total energy of the colliding molecules is not conserved. These the energy exchange processes mediated by the absorption and emission of dual microwaves lead to inelastic scattering and heating. The above analyses clearly indicate that the correct results for scatterings can only be obtained within the framework of the Floquet theory.

The experimentally more relevant quantities are the elastic and inelastic scattering rates, i.e., $\beta_{01}^{(\mathrm{el})}=v_{01}\sigma_{01}^{(\mathrm{el})}$ and $\beta_{01}^{(\mathrm{inel})}=v_{01}\sigma_{01}^{(\mathrm{inel})}$, where $v_{01}=2 \hbar k_{01}/M$ is the relative velocity. In addition, the ratio of the elastic to inelastic scattering rates, $\gamma =\beta_{01}^{(\mathrm{el})}/\beta_{01}^{(\mathrm{inel})}$ (the so-called good-to-bad collision ratio), is of particular importance for characterizing the efficiency of the evaporative cooling. Finally, from the $K$ matrix, we may compute for small $k_{01}$ the scattering length matrix according to $\mathbf{a}^{(m_0)}=-\mathbf{K}^{(m_0)}/k_{01}$, whose element $a_{n\nu l,n_0\nu_0 l_0}^{(m_0)}$ is the scattering length from the incident channel $(n_0\nu_0l_0m_0)$ to the outgoing channel $(n\nu lm_0)$. In particular, the $s$-wave scattering length for MSPMs is $a_{010,010}^{(0)}$. 

\subsection{Effective interaction between MSPMs in the Floquet theory}\label{EP}
Since the microwave-shielded $|+\rangle$ state has a sufficiently long lifetime in experiments, a molecular gas prepared in the $|+\rangle$ state represents an important platform for studying the many-body physics. As a critical ingredient for describing the molecular gases, we shall derive a time-independent effective intermolecular potential within the framework of the Floquet theory. To this end, we first note that the internal-state dynamics is much faster than the center-of-mass motion of the molecules, which allows us to employ the Born-Oppenheimer (BO) approximation. Then, for a given ${\boldsymbol r}$, we diagonalize, in the Floquet space, the potential matrix ${\mathbf V}$ (i.e., the Floquet Hamiltonian $\mathbf{H}$ with kinetic energy being neglected), which gives rise to
\begin{align}
{\mathbf V}({\boldsymbol r})\left\vert V_{n,\nu }^{(\mathrm{ad})}({\boldsymbol r})\right\rangle =V_{n,\nu }^{(\mathrm{ad})}({\boldsymbol r})\left\vert V_{n,\nu }^{(\mathrm{ad})}({\boldsymbol r})\right\rangle,  \label{V}
\end{align}
where $|V_{n,\nu}^{(\mathrm{ad})}({\boldsymbol r})\rangle$ is the eigenstate and $V_{n,\nu}^{(\mathrm{ad})}({\boldsymbol r})$ is the eigenenergy. Physically, $|V_{n,\nu}^{(\mathrm{ad})}({\boldsymbol r})\rangle$ is the state adiabatically connects to the asymptotical state $|n,\nu\rangle$, i.e., the two-molecule state $|\nu\rangle$ in the $n$-th Floquet sector. In particular, we focus on the eigenstate adiabatically connecting to $\left\vert ++\right\rangle$ with $n=0$ as $r\rightarrow \infty$. The corresponding eigenenergy $V_{0,1}^{(\mathrm{ad})}({\boldsymbol r})$ is then the effective potential between two microwave-shielded molecules. 

Alternatively, we may analytically derive an highly accurate effective potential, $V_{\mathrm{eff}}({\boldsymbol r})$, through second-order perturbation theory. For this purpose, we note that the first-order correction to the energy of the $\left\vert ++\right\rangle$ state in the sector $n=0$ is
\begin{equation}
\langle++|V_{\rm dd}({\boldsymbol r})|++\rangle= \lbrack \mathcal{V}_{0}({\boldsymbol r})]_{11}=\frac{C_{3}}{r^{3}}(3\cos
^{2}\theta -1),
\end{equation}
where
\begin{align}
C_{3}=\sqrt{\frac{15}{2\pi}}\frac{\eta}{48\pi}\left( 3\cos 2\beta -1\right) \cos ^{2}\alpha \sin
^{2}\alpha.
\end{align}
For fixed $\delta _{+}$, $\Omega _{+}$, and $\delta_{\pi }$, the Euler angle $\beta $ increases with $\Omega _{\pi }$. As $\Omega _{\pi }$ increases to the threshold value $\Omega _{c}$, $\beta$ reaches $\beta _{c}=\arccos (1/3)/2$, resulting in a complete cancellation of the effective DDI, i.e., $C_{3}=0$. Next, the second-order correction can be formally expressed as
\begin{align}
\left.\sum_{s,\nu}\right.^{\prime}\frac{\left|\left(\mathcal{V}_{s}({\boldsymbol r})\right)_{1\nu }\right|^{2}}{\mathcal{E}_{11}^{(\infty )}-\left[\mathcal{E}_{\nu\nu}^{(\infty )}-s\omega \right]},\label{2ndcorr}
\end{align}
where the primed sum excludes the term with $(s,\nu)=(0,1)$. By carefully examining the matrix elements of $\mathcal{V}_s({\boldsymbol r})$ as shown in Eq.~\eqref{Vc} and in App.~\ref{appmatele}, it turns out that, in Eq.~\eqref{2ndcorr}, only a finite number of terms contribute. Moreover interestingly, the angular dependence of the second-order corrections must be of the forms: $\left\vert Y_{20}(\hat{\boldsymbol r})\right\vert ^{2}$, $\left\vert Y_{21}(\hat{\boldsymbol r})\right\vert ^{2}$, and $\left\vert Y_{22}(\hat{\boldsymbol r})\right\vert ^{2}$. After collecting all terms contributing to the second-order correction, the effective potential can now be approximated as
\begin{align}
V_{\mathrm{eff}}({\boldsymbol r})&=\frac{C_{6}}{r^{6}}\left[ \sin ^{4}\theta +\frac{w_{1}}{w_{2}}\sin ^{2}\theta \cos ^{2}\theta +\frac{w_{0}}{w_{2}}(3\cos ^{2}\theta-1)^{2}\right]\nonumber\\
&\quad+\frac{C_{3}}{r^{3}}\left( 3\cos ^{2}\theta
-1\right),  \label{Veff}
\end{align}
where $C_{6}=15\eta^{2}w_2/(32\pi)$ and $w_{0}$, $w_1$, and $w_2$ (see App.~\ref{appeffw} for their analytical expressions) measure the relative contributions from the $\left\vert Y_{20}(\hat{\boldsymbol r})\right\vert ^{2}$, $\left\vert Y_{21}(\hat{\boldsymbol r})\right\vert ^{2}$, and $\left\vert Y_{22}(\hat{\boldsymbol r})\right\vert ^{2}$ terms, respectively. It is instructive to consider the limiting case with the $\pi$-polarized microwave being switched off. In fact, in the limit $\Omega _{\pi }\rightarrow 0$, it can then be analytically shown that $C_3\rightarrow d^2/[48\pi\epsilon_0(1+\delta_+^2)]$, $C_6\rightarrow d^4/[128\pi^2\epsilon_0^2(1+\delta_+^2)^{3/2}]$, $w_{1}/w_{2}\rightarrow 2$ and $w_0/w_2$ becomes negligibly small, which leads to exactly the effective potential derived in Ref.~\cite{Deng2023}.

\subsection{A single-channel model for MSPMs}
Armed with the effective potential, the single-channel model for the relative motion of two MSPMs is then governed by the Hamiltonian
\begin{align}
H_{\rm eff}=\frac{{\boldsymbol p}^2}{M}+V_{0,1}^{(\mathrm{ad})}({\boldsymbol r})\approx\frac{{\boldsymbol p}^2}{M}+V_{\rm eff}({\boldsymbol r}).
\end{align}
Now, for the scatterings of two MSPMs, we solve the Schr\"odinger equation 
\begin{align}
H_{\rm eff}\psi({\boldsymbol r})=\frac{\hbar^2 k_{01}^2}{M}\psi({\boldsymbol r}), 
\end{align}
where $k_{01}$ is the incident momentum. Making use of the partial-wave expansion, $\psi({\boldsymbol r})=\sum_{lm}r^{-1}\phi_{lm}(r)Y_{lm}(\hat{\boldsymbol r})$, the radial wave functions satisfy
\begin{align}
-\frac{1}{M}&\left[\frac{d^2}{dr^2}-\frac{l(l+1)}{r^2}\right]\phi_{lm}(r)\nonumber\\
&+\sum_{l'm'}[V_{{\rm eff}}(r)]_{ll'}\delta_{mm'}\phi_{l'm'}(r)=\frac{k_{01}^2}{M}\phi_{lm}(r),\label{radiawfeff}
\end{align}
where the interaction matrix elements,
\begin{align}
[V_{\rm eff}(r)]_{ll'}=\int d\hat{\boldsymbol r}Y_{lm}^{\ast}(\hat{\boldsymbol r})V_{\rm eff}({\boldsymbol r})Y_{l^{\prime }m}(\hat{\boldsymbol r}),\label{mateleveff}
\end{align}
are independent of the magnetic quantum number $m$. In analogy to the multichannel case, we numerically solve Eqs.~\eqref{radiawfeff} and compute the low-energy scattering length $a_{l,l_0}^{(m_0)}$ for the scattering from incident channel $(l_0m_0)$ to the outgoing one $(lm_0)$. Particularly, we focus on the $s$-wave scattering length $a_{0,0}^{(0)}$. 

\section{Results}\label{secresu}
In this section, we study the two- and many-body physics of MSPMs. Specifically, we shall first explore the properties of the effective potential and then calibrate it using the adiabatic potential. Next, we turn to study the low-energy scatterings of two MSPMs via both single- and multi-channel models, which further validates the effective potential. We shall also calculate the elastic and inelastic scattering rates and their ratio. These data are of particular importance in experiments. Finally, we study the ground-state properties of an ultracold gas of MSPMs.

Before presenting the results, let us specify the control parameters used in calculations. Generally, our system is characterized by the permanent dipole moment $d$, the Rabi frequencies $\Omega_+$ and the detuning $\delta_+$ of the $\sigma^+$-polarized microwave, and the Rabi frequency $\Omega_\pi$ and the detuning $\delta_\pi$ of the $\pi$-polarized microwaves. To simplify the scenario, we shall follow the the experiments by considering the bosonic NaCs molecule which possesses a permanent dipole moment of $d=4.75\, {\rm Debye}$. Moreover, the Rabi frequency and detuning of the $\sigma^+$-polarized microwave are fixed at $\Omega _{+}=2\pi \times 7.9\,\mathrm{MHz}$ and $\delta _{+}=-2\pi \times 8\,\mathrm{MHz}$, respectively. Unless otherwise specified, the detuning of the $\pi$-polarized microwave is fixed $\delta_{\pi}=-2\pi \times 10\mathrm{MHz}$. Finally, we take the Rabi frequency of the $\pi$-polarized microwave $\Omega_\pi$ as the free parameter.

\subsection{Effective potential}
\begin{figure}[tbp]
\includegraphics[width=0.95\linewidth]{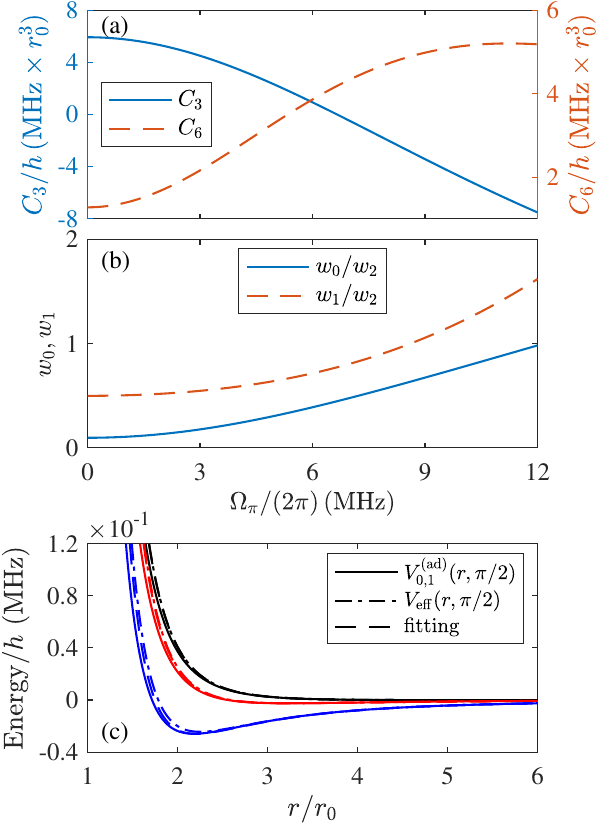}
\caption{Effective potential. (a) $C_3$ and $C_6$ as functions of $\Omega_\pi$. (b) $w_0$ and $w_1$ as functions of $\Omega_\pi$. (c) Comparison of $V_{\rm eff}({\boldsymbol r})$ with $V_{0,1}^{(\mathrm{ad})}({\boldsymbol r})$ at $\theta=\pi/2$ for $\Omega_\pi/(2\pi)=6.5$, $5.9$, and $4\,{\rm MHz}$ (for three sets of curves in descending order). The dashed lines are obtained by numerically fitting $V_{0,1}^{({\rm ad})}({\boldsymbol r})$ using Eq.~\eqref{Veff}.}
\label{potpara}
\end{figure}

As shown in Eq.~\eqref{Veff}, the effective potential is completely specified by the parameters characterizing the strength, i.e, $C_3$ and $C_6$, and the parameters characterizing the anisotropy of the shielding core, i.e., $w_0/w_2$ and $w_1/w_2$. Figure~\ref{potpara}(a) plots the $\Omega_\pi$ dependence of $C_3$ and $C_6$. As can be seen, $C_3$ is a monotonically decreasing function of $\Omega_\pi$, indicating that the $\pi$-polarized microwave indeed lowers the attraction of the negated dipolar interaction on the $xy$ plane. In particular, the dipolar interaction is completely canceled out at $\Omega_\pi=\Omega_\pi^{(c)}$ ($\approx2\pi\times 6.65\,{\rm MHz}$). Further increasing of $\Omega_\pi$ then leads to a normal dipolar interaction ($C_3<0$) that is repulsive on the $xy$ plane and attractive along the $z$ axis. On the contrary, $C_6$ is a monotonically increasing function of $\Omega_\pi$ and, as a result of the second-order perturbation, the value of $C_6$ is always positive, which clearly indicates that, similar to the $\sigma^+$-polarized microwave, the $\pi$-polarized microwave also gives rise to the shielding.

In Fig.~\ref{potpara}(b), we plot the ratios $w_0/w_2$ and $w_1/w_2$ as function of $\Omega_\pi$. At the $\Omega_\pi\rightarrow0$ limit, $w_0/w_2$ becomes negligibly small such that the effective potential~\eqref{Veff} reduces to the one derived in Ref.~\cite{Deng2023}. However, the $w_0$ term play an important role at large $\Omega_\pi$. To see this, we consider the effective potential along the $z$ axis, i.e.,
\begin{align}
V_{\rm eff}(z)=\frac{C_3}{|z|^3}+\frac{15\eta^{2}w_0}{8\pi}\frac{1}{|z|^6},
\end{align}
where only the $w_0$ term contributes to the $1/r^6$ shielding. Therefore, for negative $C_3$, the $w_0$ term is the only source providing shielding.

Finally, we compare, in Fig.~\ref{potpara}(c), the effective potential $V_{\rm eff}({\boldsymbol r})$ (dash-dotted lines) with the adiabatic potential $V_{0,1}^\mathrm{(ad)}({\boldsymbol r})$ (solid lines) on the $xy$ plane for various $\Omega_\pi$'s. Apparently, very good agreement is achieved for large $\Omega_\pi$. Although the discrepancy seems to increase as $\Omega_\pi$ is lowered, the quality of the effective potential, as shown by the dashed lines in Fig.~\ref{potpara}(c), can be improved by numerically fitting the adiabatic potential according to Eq.~\eqref{Veff}. As a result, one can always obtained the analytic expression for a highly accurate effective potential which plays an important role in studying the two- and many-body physics.

\subsection{Low-energy scatterings}
\begin{figure}[tbp]
\includegraphics[width=0.85\linewidth]{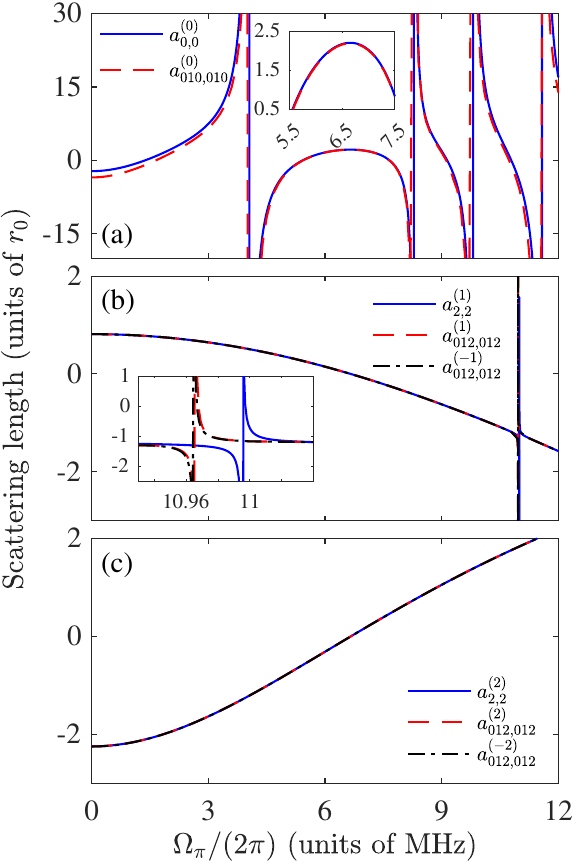}
\caption{Comparison of $a_{l_0,l_0}^{(m_0)}$ with $a_{01 l_0,01 l_0}^{(m_0)}$ for $|m_0|=0$ (a), $1$ (b), and $2$ (c). Inset in (a) is the zoom-in plot for the $s$-wave scattering length in the vicinity of $\Omega_\pi^{(c)}$ and that in (b) is the zoom-in plot of the scattering resonance with $m_0=\pm1$.}
\label{potcalib}
\end{figure}

To further justify the effective potential, here we compare, in Fig.~\ref{potcalib}, the scattering lengths computed with the single- and multi-channel models for various $m_0$. More specifically, Fig.~\ref{potcalib}(a) plots the $s$-wave scattering lengths, $a_{0,0}^{(0)}$ and $a_{010,010}^{(0)}$, as functions of $\Omega_\pi$. As can be seen, very good agreement between the single- and multi-channel calculations is achieved, which justifies the validity of the effective potential for the scattering calculations. Interestingly, the $\Omega_\pi$ dependence of the $s$-wave scattering lengths can be readily understood using the effective potential. For this purpose, we start with the cancellation point, i.e., $\Omega_\pi=\Omega_\pi^{(c)}$, where only the $C_6/r^6$ shielding potential survives. Consequently, the $s$-wave scattering length at $\Omega_\pi^{(c)}$ is always positive and, in particular, it is around $2.2r_0$ in Fig.~\ref{potcalib}(a). As $\Omega_\pi$ is deviated from $\Omega_\pi^{(c)}$, the $C_3$ term is turned on. Since the dipolar interaction is always partially attractive, the $s$-wave scattering length decreases on both sides of $\Omega_\pi^{(c)}$. Eventually, as $\Omega_\pi$ deviates sufficiently far away from $\Omega_\pi^{(c)}$, scattering resonances are experienced on both sides of $\Omega_\pi^{(c)}$, indicating the formation of the bound states. As $\Omega_\pi$ is further varied, more resonances can be encountered.

For a more comprehensive validation, we further compare the scattering lengths corresponding to the partial wave with $l_0=2$ and $|m_0|=1$ in Fig.~\ref{potcalib}(b). As can be seen, the agreement achieved here is even better than that in the $m_0=0$ case as all curves are now visually indistinguishable. Indeed, discrepancy can only be found in the zoom-in plot of the scattering resonance at $\Omega_\pi/(2\pi)\approx 11\,{\rm MHz}$ [see the inset of Fig.~\ref{potcalib}(a)]. As $|m_0|$ is further increased to $2$ in Fig.~\ref{potcalib}(c), discrepancies between $a_{2,2}^{(2)}$ and $a_{012,012}^{(\pm1)}$ completely disappear.

On the other hand, the level of the agreement achieved between $a_{l_0,l_0}^{(m_0)}$ and $a_{01 l_0,01 l_0}^{(m_0)}$ also justifies the neglecting of the induced gauge potential which combined with the adiabatic potential gives rise to a higher-order approximation, $V_{\rm tot}$, to the intermolecular potential between MSPMs. Indeed, as shown in App.~\ref{appgauge}, the leading contributions from the gauge potential scale as $1/r^8$, indicating that the gauge potential only plays role at short distance. As a result, one can hardly see the contribution from the gauge potential to the low-energy scatterings in Fig.~\ref{potcalib}. Moreover, because the gauge potential breaks the time-reversal symmetry, $V_{\rm tot}$ explicitly depends on the magnetic quantum number $m_0$ (see App.~\ref{appgauge} for details) such that the deviation of $V_{\rm tot}-V_{\rm eff}$ increases as $|m_0|$ increases. While from the scattering calculations, we see that the discrepancy between $a_{l_0,l_0}^{(m_0)}$ and $a_{01 l_0,01 l_0}^{(m_0)}$ diminishes as $m_0$ increases. The above observations can be readily understood by noting that the incident molecules with large $|m_0|$ also possess a large angular momentum ($l_0\geq |m_0|$). As a result, they experience a strong centrifugal potential, $l_0(l_0+1)/r^2$, which keeps the molecules far apart and reduces the influence of the gauge potential for low-energy scatterings. More importantly, the scattering calculations clearly indicates that the gauge potential barely plays a role in the low-energy scatterings between MSPMs. 

Here we would like to remark on a recent work~\cite{Zhang2024} in which the authors find that the gauge potential is essential for trapped polar molecules with trap frequency as high as $(2\pi)\times1\,{\rm MHz}$. From our analyses, this is not unexpected since, given such a high trap frequency, the corresponding incident energy allow two molecules to get sufficiently close to each other where the discrepancy between $V_{\rm eff}$ and $V_{\rm tot}$ is significant (see App.~\ref{appgauge}). 

\subsection{Elastic and inelastic scatterings}
\begin{figure}[tbp]
\includegraphics[width=\linewidth]{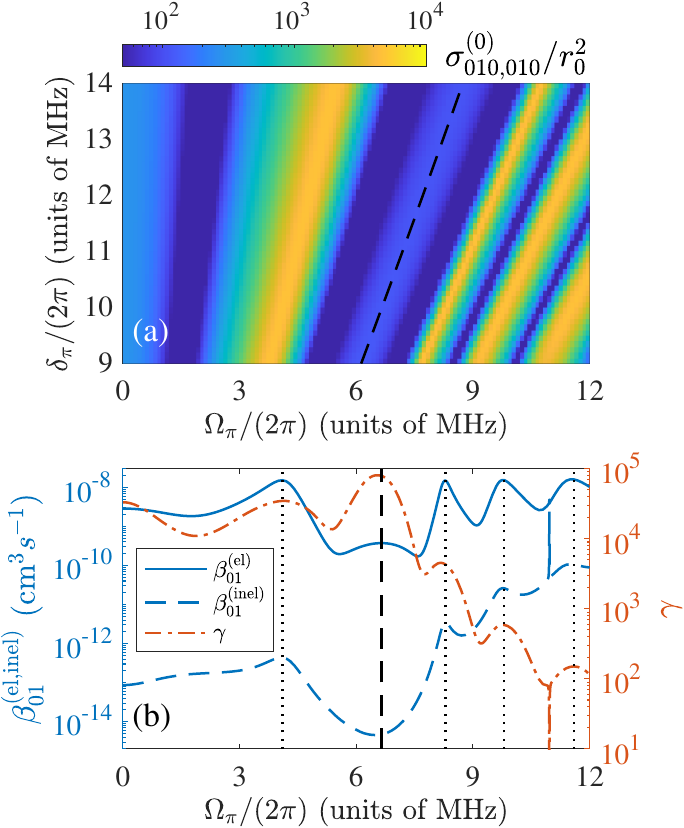}
\caption{(a) Scattering cross section $\sigma_{010,010}^{(0)}$ as a function of $\Omega_\pi$ and $\delta_\pi$. DDI is completely canceled along the black dashed line. (b) $\beta_{01}^{({\rm el})}$, $\beta_{01}^{({\rm inel})}$, and $\gamma$ versus $\Omega_\pi$ with $\delta_\pi/(2\pi)=10\,{\rm MHz}$. Here, the temperature of gas is $6$nK. The vertical dotted lines mark the positions of the resonances and the vertical dashed line marks the position of $\Omega_\pi^{(c)}$. The narrow peak at $\Omega_\pi/(2\pi)\approx 11\,{\rm MHz}$ is due to the $m_0=\pm1$ resonances.}
\label{multiscat}
\end{figure}

Here we calculate the elastic and inelastic scattering rates which are of particular importance in experiments. In Fig.~\ref{multiscat}(a), we map out the scattering cross section $\sigma_{010,010}^{(0)}$ in the $\Omega_\pi$-$\delta_\pi$ plane. The dashed line marks the cancellation Rabi frequency $\Omega_\pi^{(c)}$ along which the effective dipolar interaction is completely canceled out. As a result, molecules only experience a repulsive shielding potential during scattering and $\sigma_{010,010}^{(0)}$ along the cancellation line forms a small ridge. More interestingly, on the $\Omega_\pi$-$\delta_\pi$ plane, there also exist four other ridges with much a larger height where the scattering cross section $\sigma_{010,010}^{(0)}$ is peaked. As shown in Fig.~\ref{potcalib}(a), these ridges represent the shape resonances due to the formation of bound states.

Next, we plot, in Fig.~\ref{multiscat}(b), $\beta_{01}^{(\mathrm{el})}$, $\beta_{01}^{(\mathrm{inel})}$, and $\gamma$ as functions of $\Omega_\pi$ for $\delta_\pi/(2\pi)=10$MHz. Because the $s$ wave makes the largest contribution to the total elastic cross section, $\beta_{01}^{(\mathrm{el})}$ is roughly proportional to $[a_{010,010}^{(0)}]^2$. Consequently, $\beta_{01}^{(\mathrm{el})}$ peaks at the shape resonances and at the cancellation point. As to the inelastic scatterings, since the two-molecule bound states imply stronger couplings between the incident channel $(n,\nu)=(0,1)$ to other interaction channels, it is seen that $\beta_{01}^{(\mathrm{inel})}$ is also peaked at the shape resonances. Interestingly, $\beta_{01}^{(\mathrm{inel})}$ is minimized at the cancellation point due to the purely repulsive intermolecular potential. Finally, as to the good-to-bad collision ratio $\gamma$, since $\beta_{01}^{(\mathrm{el})}$ is at least one order of magnitude larger than $\beta_{01}^{(\mathrm{inel})}$, $\gamma$ reaches its local maxima at the shape resonances. However, the global maximum of $\gamma$ is achieved at the cancellation point where $\beta_{01}^{(\mathrm{inel})}$ is minimized. We thus identify a regime where it is most efficient for performing evaporative cooling. Remarkably, this regime is in good agreement with the choice of the experiment~\cite{Will2023b}.

\subsection{Ultracold Bose gases of MSPMs}\label{secbec}
Following the scenario of the NaCs experiments, we study the properties of the condensates of MSPMs. To this end, we first write down the many-body Hamiltonian for a gas of $N$ molecules:
\begin{align}
H&=H_0+H_{\rm int},\\
H_{0}&=\int d{\boldsymbol r}\left[ \frac{1}{2M}\nabla \hat{\psi}^{\dagger }({%
{\boldsymbol r}})\nabla \hat{\psi}({\boldsymbol r})+V({\boldsymbol r})\hat{\psi}%
^{\dagger }({\boldsymbol r})\hat{\psi}({\boldsymbol r})\right],\nonumber\\
H_{\mathrm{int}}&=\frac{1}{2}\int d{\boldsymbol r}d{\boldsymbol r}^{\prime }V_{%
\mathrm{eff}}({\boldsymbol r}-{\boldsymbol r}^{\prime })\hat{\psi}^{\dagger }({\mathbf{%
r}})\hat{\psi}^{\dagger }({\boldsymbol r}^{\prime })\hat{\psi}({\boldsymbol r}%
^{\prime })\hat{\psi}({\boldsymbol r}),\nonumber
\end{align}
where $\hat{\psi}({\boldsymbol r})$ is the field operator and $V({\boldsymbol r})=M[\omega _{\perp}^{2}(x^{2}+y^{2})+\omega _{z}^{2}z^{2}]/2$ is the confining potential with $\omega _{\perp }$ and $\omega _{z}$ being the transverse and axial trap frequencies, respectively. 

Unlike atomic condensates, strong many-body correlations may develop in ultracold molecular gases due to the large shielding core of inter-molecular potential. To incorporate the many-body correlation, we adopt the variational ansatz for the $N$-particle state~\cite{Shi2018,Jin2024}
\begin{equation}
\left\vert \Psi_{N}\right\rangle =\int \prod_{i}d{\boldsymbol r}_i\prod_{i<j(=1)}^{N}J({%
{\boldsymbol r}}_{i},{\boldsymbol r}_{j})\prod_{j=1}^{N}\phi _{0}({\boldsymbol r}_{j})%
\hat{\psi}^{\dagger }({\boldsymbol r}_{j})\left\vert 0\right\rangle,
\end{equation}%
where $\phi_{0}({\boldsymbol r})$ is the normalized single-particle wavefunction and $J({\boldsymbol r}_{i},{\boldsymbol r}_{j})$ $[=J({\boldsymbol r}_{j},{\boldsymbol r}_{i})]$ is the Jastrow correlation factor. Both $\phi_{0}({\boldsymbol r})$ and $J({\boldsymbol r}_{i},{\boldsymbol r}_{j})$ are the variational parameters which can be determined by minimizing the total energy $E[\phi_0,J]=\langle\Psi_N|\hat H|\Psi_N\rangle$ calculated via the cluster expansion~\cite{cluster1958,Jin2024}. In numerical calculations, it is more convenient to replace $\phi_0({\boldsymbol r})$ by $\sqrt{n({\boldsymbol r})}$ as the variational parameter, where $n({\boldsymbol r})=\left\langle \Psi_N \right\vert \hat{\psi}^{\dagger}({\boldsymbol r})\hat{\psi}({\boldsymbol r})\left\vert \Psi_N \right\rangle$ is the total density.

For the results of the molecular condensates presented below, we fix the axial trap frequency at $\omega_z/(2\pi)=58\,{\rm Hz}$ and the number of molecules at $N=100$. We also focus on two typical Rabi frequencies $\Omega_\pi/(2\pi)=5.9$ and $6.5\,{\rm MHz}$ which, as shall be shown, give rise to the self-bound and expanding states, respectively. The radial harmonic trap is switched on only for the expanding states and the frequency is set to $\omega_\perp/(2\pi)=33.6\,{\rm Hz}$. Finally, for convenience, we use $\varepsilon_0\equiv \hbar^2/(M r_0^2)$ as the unit for energy.

\begin{figure}[tbp]
\centering	
\includegraphics[width=.8\columnwidth]{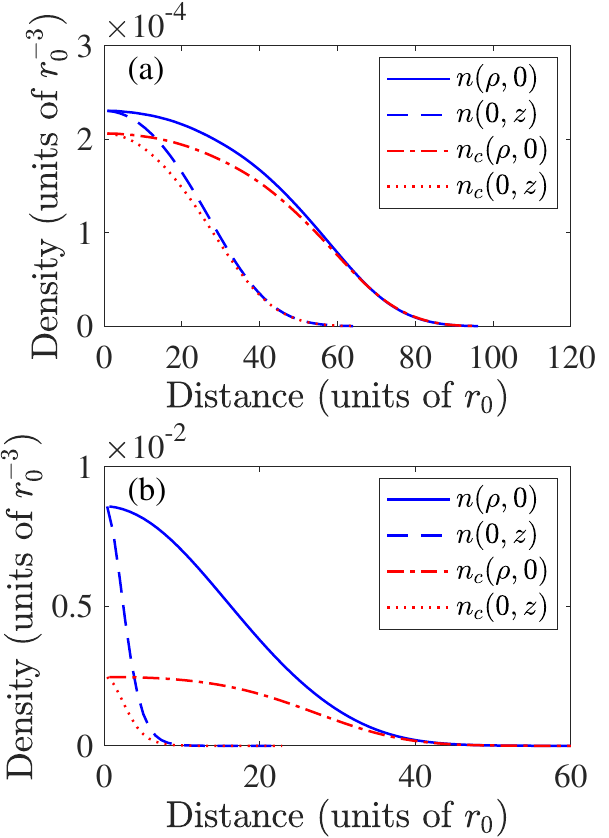}
\caption{Total and condensate densities along the radial and the $z$ directions for  $\Omega_\pi/(2\pi)=6.5$ (a) and $5.9$ MHz (b).}
\label{density}
\end{figure}

In Fig.~\ref{density}(a) and (b), we plot the total and condensate densities for the expanding and self-bound states, respectively. Here the condensate density can be obtained by diagonalizing the first-order correlation function, i.e., $G_1({\boldsymbol r},{\boldsymbol r}')=\left\langle \Psi_N \right\vert \hat{\psi}^{\dagger}({\boldsymbol r}')\hat{\psi}({\boldsymbol r})\left\vert \Psi_N \right\rangle=\sum_{\ell\geq0}N_{\ell}\bar{\varphi}_{\ell}({\boldsymbol r})\bar{\varphi}_{\ell}^{\ast }({\boldsymbol r}^{\prime })$, where $N_\ell$ (sorted in descending order) is the occupation number in the normalized mode $\bar\varphi_\ell({\boldsymbol r})$. Then $N_0$ is the number of molecules in the condensation, $n_c({\boldsymbol r})=N_0| \bar\varphi_0({\boldsymbol r})|^2$ is the corresponding condensate density and $f_c=N_0/N$ is the condensate fraction.

For the expanding state, the radial and axial widths of the total density are, respectively, $\sigma _{\rho }=43.2r_{0}$ and $\sigma_{z}=25.3r_{0}$, which, due to the repulsive nature of the interaction, are larger than the corresponding harmonic oscillator widths $a_{\perp}=26.2r_{0}$ and $a_{z}=19.8r_{0}$ of the trap. Here the radial and axial widths of the density are defined according to $\sigma_{\rho}=\left[2\pi N^{-1}\int d\rho dz\rho ^{3}n({\boldsymbol r})\right] ^{1/2}$ and $\sigma_{z}=\left[4\pi N^{-1}\int d\rho dz\rho z^{2}n({\boldsymbol r})\right] ^{1/2}$, respectively. Given that the peak density of the gas is $n_p=1.5\times 10^{12}\mathrm{cm^{-3}}$ and the scattering length is $a_s=2.16\times 10^3a_0$, it can be estimated that $na_{s}^{3}=2.3\times 10^{-3}$ ($\ll 1$), indicating that the gas is in the weak interacting regime. As a result, the condensation fraction can be as high as $f_c=0.94$.

For the self-bound state, due to the stronger attractive DDI in the $xy$ plane, the widths of the gas dramatically reduces to $\sigma_{\rho }=22.8r_0$ and $\sigma_{z}=5.1r_0$ even in the absence of the external trap. More interestingly, by examining the aspect ratios $\sigma_\rho/\sigma_z$, it is found that the self-bound state is significantly flattened radially. This phenomenon is quite common in dipolar quantum gases. In fact, to lower the interaction energy, dipolar gases always get stretched along the attractive direction of the dipolar interaction~\cite{Yi2000,Yi2001}. Moreover, as the peak density, $n_p=5.6\times 10^{13}\,\mathrm{cm^{-3}}$, becomes dramatically higher than that of the expanding state, the condensate fraction reduces to $f_{c}=0.53$, indicating that the self-bound state corresponds to a more strongly correlated gas. 

\begin{figure}[tbp]
\centering	
\includegraphics[width=.8\columnwidth]{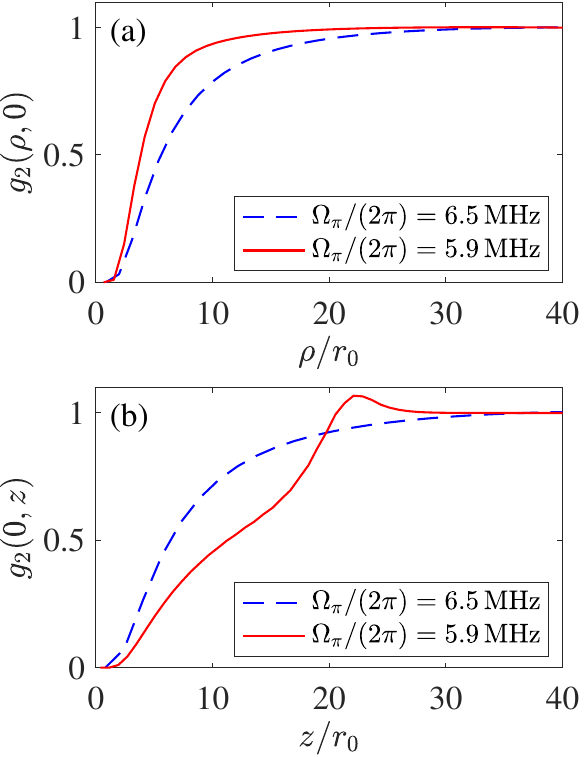}
\caption{Normalized second-order correlation function $g_2(\rho,0)$ (a) and $g_2(0,z)$ (b) for $\Omega_\pi/(2\pi)=5.9$ (solid lines) and $6.5\,{\rm MHz}$ (dashed lines).}
\label{g2rhoz}
\end{figure}

To measure many-body correlation, we calculate the normalized second-order correlation function
\begin{align}
g_{2}({\boldsymbol r},{\boldsymbol r}')=\frac{G_2({\boldsymbol r},{\boldsymbol r}') }{n({\boldsymbol r})n({\boldsymbol r}')},
\end{align}
where $G_2({\boldsymbol r},{\boldsymbol r}')=\left\langle \Psi\right\vert \hat{\psi}^{\dagger }({\boldsymbol r})\hat{\psi}^{\dagger }({\boldsymbol r}')\hat{\psi}({\boldsymbol r}')\hat{\psi}({\mathbf{r}})\left\vert \Psi \right\rangle$. Figures~\ref{g2rhoz}(a) and (b) plot, for two different $\Omega_\pi$'s, $g_2(\rho,0)$ and $g_2(0,z)$, respectively, where $g_2(\rho,z)\equiv g_{2}({\boldsymbol r},{\boldsymbol r}'=0)$ because of the cylindrical symmetry. Generally, $g_{2}$ vanishes at short distances due to the strong shielding potential and approaches unity at large distance. This anti-bunching behavior reveals that each molecule is surrounded by holes at short distance. Consequently, the coherence can only be established as the distance between molecules is sufficiently large, which results a reduced the condensate fraction. Next, to understand the behaviors of $g_2$ presented in Fig.~\ref{g2rhoz}, we recall that the dipolar interaction is much stronger with $\Omega_\pi/(2\pi)=5.9\,{\rm MHz}$ and, in particular, the long-range interaction is nearly canceled for $\Omega_\pi/(2\pi)=6.5\,{\rm MHz}$. Then, it is natural to see [Fig.~\ref{g2rhoz}(a)] that $g_2(\rho,0)$ approaches unity more rapidly for smaller $\Omega_\pi$ since it corresponds to a stronger attraction along the radial direction at large distance. Moreover, although the interactions along the $z$ axis are purely repulsive in both cases, smaller $\Omega_\pi$ has a stronger interaction strength and a longer interaction range. Consequently, $g_2(0,z)$ with smaller $\Omega_\pi$ approaches unity more slowly [Fig.~\ref{g2rhoz}(b)]. More interestingly, we even observe the Friedal oscillation at $\rho\approx 20r_0$ where the density of the gases becomes negligibly small. This is clearly a demonstration of the strong correlation.

\begin{figure}[tbp]
\centering	
\includegraphics[width=1\columnwidth]{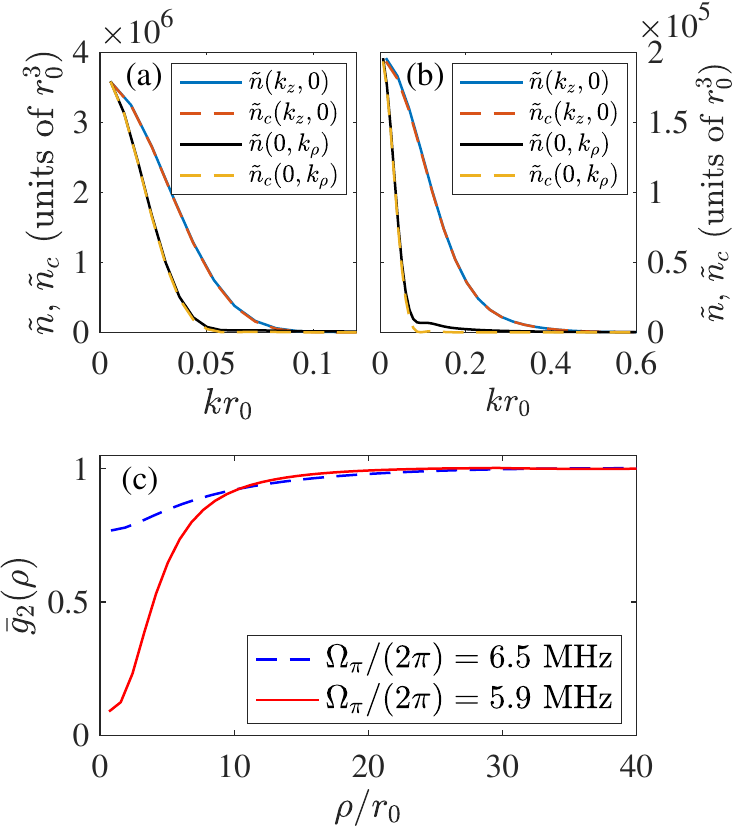}
\caption{Momentum distributions of the total molecular gas (solid line) and the condensed part (dashed line) along the radial $k_\rho$ and the $k_z$ directions for  $\Omega_\pi/(2\pi)=6.5$ (a) and $5.9$ MHz (b). (c) The integrated second-order correlation functions.}
\label{momdis}
\end{figure}

As to the experimental detection of the molecular condensates, we explore the momentum distribution of the gas. For all and condensed molecules, the momentum distributions are, respectively,
\begin{eqnarray}
\tilde{n}(\mathbf{k}) &=&\int \frac{d^{3}{\boldsymbol r}d^{3}{\boldsymbol r}^{\prime }
}{(2\pi )^{3}}e^{i\mathbf{k}\cdot ({\boldsymbol r}^{\prime }-{\boldsymbol r})}G_{1}(
{\boldsymbol r},{\boldsymbol r}^{\prime }),  \\
\tilde{n}_{c}(\mathbf{k}) &=&\frac{N_{0}}{(2\pi )^{3}}\left\vert \int d^{3}
{\boldsymbol r}e^{-i\mathbf{k}\cdot {\boldsymbol r}}\bar{\varphi}_{0}({\boldsymbol r}
)\right\vert ^{2},
\end{eqnarray}
which can be measured in the TOF experiment.

In Fig.~\ref{momdis}(a) and (b), we plot the momentum distributions $\tilde{n}(\mathbf{k})$ and $\tilde{n}_{c}(\mathbf{k})$ of total and condensed molecules for $\Omega_{\pi }/(2\pi )=6.5$ and $5.9$MHz. For larger $\Omega _{\pi }/(2\pi )$, $\tilde{n}(\mathbf{k})$ and $\tilde{n}_{c}(\mathbf{k})$ are nearly identical along all directions due to the large condensate fraction; while for smaller $\Omega_{\pi}$, discrepancy emerges between two distributions on the $xy$ plane. In particular, it is seen that the condensed molecules are dominant for small $k$ and the uncondensed molecules occupy the large $k$ region. Consequently, $\tilde n(k)$ generally exhibits a bimodal density distribution even at zero temperature.

Finally, we consider the integrated second-order correlation function
\begin{align}
\bar g_2({\boldsymbol\rho},{\boldsymbol\rho}')=\frac{\int dzdz'G_2({\boldsymbol r},{\boldsymbol r}')}{\int dzdz'n({\boldsymbol r})n({\boldsymbol r}')},\label{intg2}
\end{align}
which is measurable in experiments. Particularly, we are interested in $\bar g_2(\rho)\equiv \bar g_2({\boldsymbol\rho},0)$ which possesses an axial symmetry. Figure~\ref{momdis}(c) shows $\bar g_2(\rho)$ for two different $\Omega_\pi$'s. As can be seen, although antibunching behaviors are observed, $\bar g_2(0)$ remains finite, which is in striking contrast to the three-dimensional $g_2(0,0)$ presented in Fig.~\ref{g2rhoz}. By carefully examine Eq.~\eqref{intg2}, the nonzero $\bar g_2(0)$ is clearly contributed by the second-order correlation of the form $G_2({\boldsymbol\rho},z;{\boldsymbol\rho},z')$. Moreover, due to the stronger repulsion at short distance and the smaller width along the $z$ axis, the gas with smaller $\Omega_\pi$ experiences a stronger antibunching.

\section{Conclusion}\label{secconcl}
In conclusion, we have studied the two- and many-body physics rotating polar molecules subjected to a $\sigma^+$- and a $\pi$-polarized microwaves. With such dual microwave configuration, the two-body interaction is periodically dependent on time, although the single-molecule Hamiltonian is time-independent under a suitable rotating frame. Consequently, we have to treat the two-body physics within the framework of the Floquet theory. In particular, we compute (in)elastic scattering rates through the multi-channel scattering calculations, which allows us to identify a regimes where the elastic-to-inelastic scattering ratio is optimized for evaporative cooling, in good agreement with the choice of the experiments. Moreover, we have analytically derived an effective potential between two MSPMs which, when applied to scattering problems, provides intuitive insights for the multi-channel scattering calculations. From the many-body perspectives, our studies on the ground-state properties position MSPMs as a highly tunable platform for exploring novel many-body phenomena, transcending the paradigms of dilute atomic BECs~\cite{BECreview2002,Pethick_Smith_2008} and superfluid helium~\cite{Michels1938,Michels1939}.

\begin{acknowledgments}
This work was supported by the NSFC (Grants No. 12135018, No.12047503, and No. 12274331), by National Key Research and Development Program of China (Grant No. 2021YFA0718304), and by CAS Project for Young Scientists in Basic Research (Grant No. YSBR-057).
\end{acknowledgments}

\bibliography{ref_dMolecule}
\clearpage

\widetext

\appendix

\section{Single-molecule eigenstates and two-body interactions}\label{appsingle}

In the basis $\left\{ \left\vert 0,0\right\rangle ,\left\vert
1,1\right\rangle ,\left\vert 1,0\right\rangle ,\left\vert 1,-1\right\rangle
\right\} $, the single-molecule Hamiltonian reads%
\begin{equation}
\hat{h}_{\mathrm{in}}=%
\begin{pmatrix}
0 & \frac{\Omega _{+}}{2} & \frac{\Omega _{\pi }}{2} & 0 \\
\frac{\Omega _{+}}{2} & \delta _{+} & 0 & 0 \\
\frac{\Omega _{\pi }}{2} & 0 & \delta _{\pi } & 0 \\
0 & 0 & 0 & \delta _{+}%
\end{pmatrix}%
,
\end{equation}%
which can be diagonalized via the transformation%
\begin{equation}
U_{1}=%
\begin{pmatrix}
\cos \alpha & 0 & -\sin \alpha \cos \gamma & \sin \alpha \sin \gamma \\
\sin \alpha \cos \beta & 0 & \cos \alpha \cos \beta \cos \gamma -\sin \beta
\sin \gamma & -\cos \alpha \cos \beta \sin \gamma -\sin \beta \cos \gamma \\
\sin \alpha \sin \beta & 0 & \cos \alpha \sin \beta \cos \gamma +\cos \beta
\sin \gamma & -\cos \alpha \sin \beta \sin \gamma +\cos \beta \cos \gamma \\
0 & 1 & 0 & 0%
\end{pmatrix}%
\end{equation}
parametrized by the three Euler angles $\alpha $, $\beta $, and $\gamma $.
The columns of $U_{1}$ are the eigenvectors of $\hat{h}_{\mathrm{in}}$
which, from left to right, are denoted as $\left\vert +\right\rangle ^{(\pi
)}$, $\left\vert -1\right\rangle ^{(\pi )}=\left\vert 1,-1\right\rangle $, $%
\left\vert -\right\rangle ^{(\pi )}$, and $\left\vert 0\right\rangle ^{(\pi
)}$. The corresponding eigenenergies are $E_{+}^{(\pi )}$, $E_{-1}^{(\pi
)}=\delta _{+}$, $E_{-}^{(\pi )}$, and $E_{0}^{(\pi )}$, respectively.

\section{Matrix elements of $\Sigma_{2,m}^{(s)}$ in Eq.~\eqref{2bodyint}}\label{appmatele}
Here we explicitly list the nonzero elements of the matrix $\Sigma_{2,m}^{(s)}$. The nonzero elements of $\Sigma_{2,0}^{(0)}$ are
\begin{align}
\lbrack \Sigma _{2,0}^{(0)}]_{11} &=\frac{1-3\cos 2\beta }{4\sqrt{6}\pi }%
\sin ^{2}\alpha \cos ^{2}\alpha ,  \notag \\
\lbrack \Sigma _{2,0}^{(0)}]_{31} &=[\Sigma _{2,0}^{(0)}]_{13}=\frac{\sin
\alpha \cos \alpha }{8\sqrt{3}\pi }[\cos 2\alpha (1-3\cos 2\beta )\cos
\gamma +3\cos \alpha \sin 2\beta \sin \gamma ],  \notag \\
\lbrack \Sigma _{2,0}^{(0)}]_{71} &=[\Sigma _{2,0}^{(0)}]_{17}=\frac{\sin
^{2}\alpha \cos \alpha \cos \gamma }{4\sqrt{6}\pi }[\cos \alpha (3\cos
2\beta -1)\cos \gamma -3\sin 2\beta \sin \gamma ],  \notag \\
\lbrack \Sigma _{2,0}^{(0)}]_{22} &=-\frac{\cos ^{2}\alpha }{4\sqrt{6}\pi }%
,[\Sigma _{2,0}^{(0)}]_{52}=[\Sigma _{2,0}^{(0)}]_{25}=\frac{\sin \alpha
\cos \alpha \cos \gamma }{4\sqrt{6}\pi },  \notag \\
\lbrack \Sigma _{2,0}^{(0)}]_{33} &=\frac{1}{8\sqrt{6}\pi }\{(\cos
^{4}\alpha +\sin ^{4}\alpha )(1-3\cos 2\beta )\cos ^{2}\gamma +3\cos \alpha
\cos 2\alpha \sin 2\beta \sin 2\gamma  \notag \\
&\quad+\cos ^{2}\alpha \lbrack 2\sin ^{2}\alpha (3\cos 2\beta -1)\cos ^{2}\gamma
+(1+3\cos 2\beta )\sin ^{2}\gamma ]\},  \notag \\
\lbrack \Sigma _{2,0}^{(0)}]_{73} &=[\Sigma _{2,0}^{(0)}]_{37}=\frac{\sin
\alpha \cos \gamma }{8\sqrt{3}\pi }\{\cos \alpha \lbrack \cos 2\alpha (3\cos
2\beta -1)\cos ^{2}\gamma -(1+3\cos 2\beta )\sin ^{2}\gamma ]-\frac{3}{4}%
(1+3\cos 2\alpha )\sin 2\beta \sin 2\gamma \},  \notag \\
\lbrack \Sigma _{2,0}^{(0)}]_{44} &=\frac{1}{8\sqrt{6}\pi }[\cos ^{2}\alpha
(1+3\cos 2\beta )\cos ^{2}\gamma +\cos ^{2}2\alpha (1-3\cos 2\beta )\sin
^{2}\gamma -3\cos \alpha \cos 2\alpha \sin 2\beta \sin 2\gamma ],  \notag \\
\lbrack \Sigma _{2,0}^{(0)}]_{84} &=[\Sigma _{2,0}^{(0)}]_{48}=\frac{\sin
\alpha }{4\sqrt{6}\pi }[\cos \alpha \left( \sin ^{2}\beta \cos \gamma \cos
2\gamma -2\cos ^{2}\beta \cos ^{3}\gamma \right) +4\cos ^{3}\alpha \cos
2\beta \cos \gamma \sin ^{2}\gamma  \notag \\
&\qquad\qquad\quad\;+\sin \alpha \sin \beta \sin \gamma (\sin 2\alpha \sin \beta \sin 2\gamma
-3\sin \alpha \cos \beta \cos 2\gamma )+\frac{3}{2}\cos ^{2}\alpha \sin
2\beta \sin 3\gamma ],  \notag \\
\lbrack \Sigma _{2,0}^{(0)}]_{55} &=-\frac{\sin ^{2}\alpha \cos ^{2}\gamma
}{4\sqrt{6}\pi },[\Sigma _{2,0}^{(0)}]_{66}=-\frac{\sin ^{2}\alpha \sin
^{2}\gamma }{4\sqrt{6}\pi },  \notag \\
\lbrack \Sigma _{2,0}^{(0)}]_{77} &=\frac{\sin ^{2}\alpha \cos ^{2}\gamma }{%
2\sqrt{6}\pi }[2(\cos \alpha \sin \beta \cos \gamma +\cos \beta \sin \gamma
)^{2}-(\cos \alpha \cos \beta \cos \gamma -\sin \beta \sin \gamma )^{2}],
\notag \\
\lbrack \Sigma _{2,0}^{(0)}]_{88} &=\frac{\sin ^{2}\alpha }{64\sqrt{6}\pi }%
[6+2\cos 2\alpha +6\cos 2\beta -3\cos 2(\alpha -\beta )-3\cos 2(\alpha
+\beta )  \notag \\
&\quad+\cos 4\gamma \lbrack 6(3+\cos 2\alpha )\cos 2\beta +4\sin ^{2}\alpha
]-24\cos \alpha \sin 2\beta \sin 4\gamma ],  \notag \\
\lbrack \Sigma _{2,0}^{(0)}]_{99} &=\frac{\sin ^{2}\alpha \sin ^{2}\gamma }{%
2\sqrt{6}\pi }[2(\cos \beta \cos \gamma -\cos \alpha \sin \beta \sin \gamma
)^{2}-(\cos \gamma \sin \beta +\cos \alpha \cos \beta \sin \gamma )^{2}].\nonumber
\end{align}
The nonzero elements of $\Sigma_{2,0}^{(1)}$ are
\begin{align}
\lbrack \Sigma _{2,0}^{(1)}]_{41} &=\frac{\sin \alpha }{16\sqrt{3}\pi }%
[6\cos ^{2}\alpha \sin 2\beta \cos \gamma +(\cos \alpha +\cos 3\alpha
)(3\cos 2\beta -1)\sin \gamma ],  \notag \\
\lbrack \Sigma _{2,0}^{(1)}]_{81} &=-\frac{\sin ^{2}\alpha \cos \alpha }{8%
\sqrt{3}\pi }[3\sin 2\beta \cos 2\gamma +\cos \alpha (3\cos 2\beta -1)\sin
2\gamma ],  \notag \\
\lbrack \Sigma _{2,0}^{(1)}]_{62} &=-\frac{\sin \alpha \cos \alpha \sin
\gamma }{4\sqrt{6}\pi },\nonumber\\
[\Sigma _{2,0}^{(1)}]_{65}&=\frac{\sin ^{2}\alpha
\sin \gamma \cos \gamma }{4\sqrt{6}\pi },  \notag \\
\lbrack \Sigma _{2,0}^{(1)}]_{43} &=\frac{1}{32\sqrt{6}\pi }\{6(\cos \alpha
+\cos 3\alpha )\sin 2\beta \cos 2\gamma +[\cos 2\alpha -\cos 4\alpha
+3(2+\cos 2\alpha +\cos 4\alpha )\cos 2\beta ]\sin 2\gamma \}  \notag \\
\lbrack \Sigma _{2,0}^{(1)}]_{83} &=\frac{\sin \alpha }{8\sqrt{6}\pi }%
\{3\left( \cos \gamma \cos 2\gamma \sin ^{2}\alpha -\cos ^{2}\alpha \cos
3\gamma \right) \sin 2\beta +\cos \alpha (1+3\cos 2\beta )\sin ^{3}\gamma
\notag \\
&\quad+[\cos 3\alpha -3(2\cos \alpha +\cos 3\alpha )\cos 2\beta ]\cos ^{2}\gamma
\sin \gamma \}  \notag \\
\lbrack \Sigma _{2,0}^{(1)}]_{94} &=\frac{\sin \alpha \sin \gamma }{8\sqrt{3%
}\pi }\{\frac{3}{2}\sin ^{2}\alpha \sin 2\beta \sin 2\gamma +\cos ^{3}\alpha
(1-3\cos 2\beta )\sin ^{2}\gamma -3\cos ^{2}\alpha \sin 2\beta \sin 2\gamma
\notag \\
&\quad+\cos \alpha \lbrack (1+3\cos 2\beta )\cos [\gamma ]^{2}+(3\cos 2\beta
-1)\sin ^{2}\alpha \sin ^{2}\gamma ]\}  \notag \\
\lbrack \Sigma _{2,0}^{(1)}]_{47} &=-\frac{\sin \alpha \cos \gamma }{32%
\sqrt{3}\pi }\{3[(1+3\cos 2\alpha )\cos 2\gamma +2\sin ^{2}\alpha ]\sin
2\beta +4\cos \alpha \left( 3\cos ^{2}\alpha \cos 2\beta +\sin ^{2}\alpha
\right) \sin 2\gamma \},  \notag \\
\lbrack \Sigma _{2,0}^{(1)}]_{87} &=\frac{\sin ^{2}\alpha \cos \gamma }{8%
\sqrt{3}\pi }[3\cos \alpha \sin 2\beta \cos 3\gamma +2\cos ^{2}\alpha (3\cos
2\beta -1)\sin \gamma \cos ^{2}\gamma +(1+3\cos 2\beta )\cos 2\gamma \sin
\gamma ],  \notag \\
\lbrack \Sigma _{2,0}^{(1)}]_{98} &=\frac{\sin ^{2}\alpha \sin \gamma }{32%
\sqrt{3}\pi }[-4\cos ^{3}\gamma -3\cos 2\beta (\cos \gamma +3\cos 3\gamma
)+4\cos 2\alpha (3\cos 2\beta -1)\sin ^{2}\gamma \cos \gamma +12\cos \alpha
\sin 2\beta \sin 3\gamma ].\nonumber
\end{align}
The nonzero elements of $\Sigma_{2,0}^{(2)}$ are
\begin{align}
\lbrack \Sigma _{2,0}^{(2)}]_{91} &=\frac{\cos \alpha \sin ^{2}\alpha \sin
\gamma }{4\sqrt{6}\pi }[3\sin 2\beta \cos \gamma +\cos \alpha (3\cos 2\beta
-1)\sin \gamma ],  \notag \\
\lbrack \Sigma _{2,0}^{(2)}]_{93} &=\frac{\sin \alpha \sin \gamma }{32\sqrt{%
3}\pi }\{3[(1+3\cos 2\alpha )\cos 2\gamma -2\sin ^{2}\alpha ]\sin 2\beta
+4\cos \alpha \left( 3\cos ^{2}\alpha \cos 2\beta +\sin ^{2}\alpha \right)
\sin 2\gamma \},  \notag \\
\lbrack \Sigma _{2,0}^{(2)}]_{97} &=-\frac{\sin ^{2}\alpha \sin 2\gamma }{32%
\sqrt{6}\pi }\{12\cos \alpha \sin 2\beta \cos 2\gamma +[3(3+\cos 2\alpha
)\cos 2\beta +2\sin ^{2}\alpha ]\sin 2\gamma \}.\nonumber
\end{align}%
The nonzero elements of $\Sigma_{2,1}^{(-2)}$ are
\begin{align}
\lbrack \Sigma _{2,1}^{(-2)}]_{16} &=-\frac{1}{4\pi }\sin ^{2}\alpha \cos
\alpha \sin \beta \sin \gamma ,  \notag \\
\lbrack \Sigma _{2,1}^{(-2)}]_{36} &=-\frac{\sin \alpha \sin \gamma }{4%
\sqrt{2}\pi }(\cos 2\alpha \sin \beta \cos \gamma +\cos \alpha \cos \beta
\sin \gamma ),  \notag \\
\lbrack \Sigma _{2,1}^{(-2)}]_{76} &=\frac{\sin ^{2}\alpha \sin 2\gamma }{%
8\pi }(\cos \alpha \sin \beta \cos \gamma +\cos \beta \sin \gamma ).\nonumber
\end{align}
The nonzero elements of $\Sigma_{2,1}^{(-1)}$ are
\begin{align}
\lbrack \Sigma _{2,1}^{(-1)}]_{12} &=-\frac{1}{4\pi }\cos ^{2}\alpha \sin
\alpha \sin \beta ,\nonumber\\
\lbrack \Sigma _{2,1}^{(-1)}]_{15}&=\frac{1}{4\pi }\sin^{2}\alpha \cos \alpha \sin \beta \cos \gamma ,  \notag \\
\lbrack \Sigma _{2,1}^{(-1)}]_{32} &=-\frac{1}{8\sqrt{2}\pi }[(\cos \alpha
+\cos 3\alpha )\sin \beta \cos \gamma +2\cos ^{2}\alpha \cos \beta \sin
\gamma ],  \notag \\
\lbrack \Sigma _{2,1}^{(-1)}]_{72} &=\frac{\sin \alpha \cos \alpha \cos
\gamma }{4\pi }(\cos \alpha \sin \beta \cos \gamma +\cos \beta \sin \gamma ),
\notag \\
\lbrack \Sigma _{2,1}^{(-1)}]_{35} &=\frac{\sin \alpha \cos \gamma }{4\sqrt{%
2}\pi }(\cos 2\alpha \sin \beta \cos \gamma +\cos \alpha \cos \beta \sin
\gamma ),  \notag \\
\lbrack \Sigma _{2,1}^{(-1)}]_{75} &=-\frac{\sin ^{2}\alpha \cos ^{3}\gamma
}{4\pi }(\cos \alpha \sin \beta +\cos \beta \tan \gamma ),  \notag \\
\lbrack \Sigma _{2,1}^{(-1)}]_{46} &=\frac{\sin \alpha \sin \gamma }{4\sqrt{%
2}\pi }(\cos 2\alpha \sin \beta \sin \gamma -\cos \alpha \cos \beta \cos
\gamma ),  \notag \\
\lbrack \Sigma _{2,1}^{(-1)}]_{86} &=\frac{\sin ^{2}\alpha \sin \gamma }{4%
\sqrt{2}\pi }(\cos \beta \cos 2\gamma -\cos \alpha \sin \beta \sin 2\gamma ),
\notag \\
\lbrack \Sigma _{2,1}^{(-1)}]_{19} &=\frac{\sin ^{2}\alpha \cos \alpha \cos
\beta \sin \gamma }{2\sqrt{2}\pi }(\cos \beta \cos \gamma -\cos \alpha \sin
\beta \sin \gamma ),  \notag \\
\lbrack \Sigma _{2,1}^{(-1)}]_{39} &=\frac{\sin \alpha \sin \gamma }{4\pi }%
(\cos \alpha \sin \beta \sin \gamma -\cos \beta \cos \gamma )(\cos \alpha
\sin \beta \sin \gamma -\cos 2\alpha \cos \beta \cos \gamma ),  \notag \\
\lbrack \Sigma _{2,1}^{(-1)}]_{79} &=\frac{\sin ^{2}\alpha \cos \gamma \sin
\gamma }{2\sqrt{2}\pi }(\cos \alpha \sin \beta \sin \gamma -\cos \beta \cos
\gamma )(\cos \alpha \cos \beta \cos \gamma -\sin \beta \sin \gamma ).\nonumber
\end{align}
The nonzero elements of $\Sigma_{2,1}^{(0)}$ are
\begin{align}
\lbrack \Sigma _{2,1}^{(0)}]_{42} &=\frac{1}{8\sqrt{2}\pi }[(\cos \alpha
+\cos 3\alpha )\sin \beta \sin \gamma -2\cos ^{2}\alpha \cos \beta \cos
\gamma ],  \notag \\
\lbrack \Sigma _{2,1}^{(0)}]_{82} &=\frac{\sin 2\alpha }{8\sqrt{2}\pi }%
(\cos \beta \cos 2\gamma -\cos \alpha \sin \beta \sin 2\gamma ),  \notag \\
\lbrack \Sigma _{2,1}^{(0)}]_{14} &=\frac{1}{32\pi }(8\sin \alpha \cos
^{2}\alpha \cos ^{2}\beta \cos \gamma -\sin 4\alpha \sin 2\beta \sin \gamma
),  \notag \\
\lbrack \Sigma _{2,1}^{(0)}]_{34} &=\frac{1}{32\sqrt{2}\pi }[2(\cos \alpha
+\cos 3\alpha )(1+\cos 2\beta \cos 2\gamma )-(2+\cos 2\alpha +\cos 4\alpha
)\sin 2\beta \sin 2\gamma ],  \notag \\
\lbrack \Sigma _{2,1}^{(0)}]_{74} &=-\frac{\sin \alpha \cos \gamma }{4\pi }%
(\cos \alpha \cos \beta \cos \gamma -\sin \beta \sin \gamma )(\cos \alpha
\cos \beta \cos \gamma -\cos 2\alpha \sin \beta \sin \gamma ),  \notag \\
\lbrack \Sigma _{2,1}^{(0)}]_{45} &=\frac{\sin \alpha \cos \gamma }{4\sqrt{2%
}\pi }(\cos \alpha \cos \beta \cos \gamma -\cos 2\alpha \sin \beta \sin
\gamma ),  \notag \\
\lbrack \Sigma _{2,1}^{(0)}]_{85} &=\frac{\sin ^{2}\alpha \cos \gamma }{4%
\sqrt{2}\pi }(\cos \alpha \sin \beta \sin 2\gamma -\cos \beta \cos 2\gamma ),
\notag \\
\lbrack \Sigma _{2,1}^{(0)}]_{96} &=\frac{\sin ^{2}\alpha \sin ^{3}\gamma }{%
4\pi }(\cos \alpha \sin \beta -\cos \beta \cot \gamma ),  \notag \\
\lbrack \Sigma _{2,1}^{(0)}]_{18} &=\frac{\sin ^{2}\alpha \cos \alpha \cos
\beta }{4\pi }(\cos \alpha \sin \beta \sin 2\gamma -\cos \beta \cos 2\gamma )
\notag \\
\lbrack \Sigma _{2,1}^{(0)}]_{38} &=\frac{\sin \alpha }{4\sqrt{2}\pi }(\cos
\alpha \sin \beta \sin \gamma -\cos 2\alpha \cos \beta \cos \gamma )(\cos
\beta \cos 2\gamma -\cos \alpha \sin \beta \sin 2\gamma ),  \notag \\
\lbrack \Sigma _{2,1}^{(0)}]_{78} &=\frac{\sin ^{2}\alpha \cos ^{2}\gamma }{%
4\pi }(\cos \beta \cos 2\gamma -\cos \alpha \sin \beta \sin 2\gamma )(\cos
\alpha \cos \beta -\sin \beta \tan \gamma ),  \notag \\
\lbrack \Sigma _{2,1}^{(0)}]_{49} &=\frac{\sin \alpha \sin \gamma }{4\pi }%
(\cos \alpha \sin \beta \cos \gamma +\cos 2\alpha \cos \beta \sin \gamma
)(\cos \alpha \sin \beta \sin \gamma -\cos \beta \cos \gamma ),  \notag \\
\lbrack \Sigma _{2,1}^{(0)}]_{89} &=\frac{\sin ^{2}\alpha \sin \gamma }{%
4\pi }(\cos \beta \cos \gamma -\cos \alpha \sin \beta \sin \gamma )(\sin
\beta \cos 2\gamma +\cos \alpha \cos \beta \sin 2\gamma ).\nonumber
\end{align}
The nonzero elements of $\Sigma_{2,1}^{(1)}$ are
\begin{align}
\lbrack \Sigma _{2,1}^{(1)}]_{11} &=\frac{\sin ^{2}2\alpha \sin 2\beta }{16%
\sqrt{2}\pi },\nonumber\\
\lbrack \Sigma _{2,1}^{(1)}]_{31}&=\frac{1}{32\pi }(\sin 4\alpha \sin
2\beta \cos \gamma -8\sin \alpha \cos ^{2}\alpha \sin ^{2}\beta \sin \gamma),  \notag \\
\lbrack \Sigma _{2,1}^{(1)}]_{71} &=\frac{\sin ^{2}\alpha \cos \alpha \sin
\beta \cos \gamma }{2\sqrt{2}\pi }(\sin \beta \sin \gamma -\cos \alpha \cos
\beta \cos \gamma ),  \notag \\
\lbrack \Sigma _{2,1}^{(1)}]_{92} &=\frac{\sin 2\alpha \sin \gamma }{8\pi }%
(\cos \alpha \sin \beta \sin \gamma -\cos \beta \cos \gamma ),  \notag \\
\lbrack \Sigma _{2,1}^{(1)}]_{13} &=\frac{1}{32\pi }(\sin 4\alpha \sin
2\beta \cos \gamma +8\sin \alpha \cos ^{2}\alpha \cos ^{2}\beta \sin \gamma),  \notag \\
\lbrack \Sigma _{2,1}^{(1)}]_{33} &=\frac{1}{16\sqrt{2}\pi }[\sin 2\beta
\left( \cos 4\alpha \cos ^{2}\gamma +\cos 2\gamma -\cos 2\alpha \sin
^{2}\gamma \right) +(\cos \alpha +\cos 3\alpha )\cos 2\beta \sin 2\gamma ],
\notag \\
\lbrack \Sigma _{2,1}^{(1)}]_{73} &=-\frac{\sin \alpha \cos \gamma }{4\pi }%
(\cos 2\alpha \cos \gamma \sin \beta +\cos \alpha \cos \beta \sin \gamma
)(\cos \alpha \cos \beta \cos \gamma -\sin \beta \sin \gamma ),  \notag \\
\lbrack \Sigma _{2,1}^{(1)}]_{44} &=-\frac{1}{32\sqrt{2}\pi }\{[\cos
2\alpha -\cos 4\alpha +(2+\cos 2\alpha +\cos 4\alpha )\cos 2\gamma ]\sin
2\beta +2(\cos \alpha +\cos 3\alpha )\cos 2\beta \sin 2\gamma \},  \notag \\
\lbrack \Sigma _{2,1}^{(1)}]_{84} &=\frac{\sin \alpha }{4\sqrt{2}\pi }(\cos
\alpha \cos \beta \cos \gamma -\cos 2\alpha \sin \beta \sin \gamma )(\sin
\beta \cos 2\gamma +\cos \alpha \cos \beta \sin 2\gamma ),  \notag \\
\lbrack \Sigma _{2,1}^{(1)}]_{95} &=\frac{\sin ^{2}\alpha \sin 2\gamma }{%
8\pi }(\cos \beta \cos \gamma -\cos \alpha \sin \beta \sin \gamma ),  \notag
\\
\lbrack \Sigma _{2,1}^{(1)}]_{17} &=-\frac{\sin ^{2}\alpha \cos \alpha \cos
\beta \cos \gamma }{2\sqrt{2}\pi }(\cos \alpha \sin \beta \cos \gamma +\cos
\beta \sin \gamma ),  \notag \\
\lbrack \Sigma _{2,1}^{(1)}]_{37} &=\frac{\sin \alpha \cos \gamma }{4\pi }%
(\cos \alpha \sin \beta \cos \gamma +\cos \beta \sin \gamma )(\cos \alpha
\sin \beta \sin \gamma -\cos 2\alpha \cos \beta \cos \gamma ),  \notag \\
\lbrack \Sigma _{2,1}^{(1)}]_{77} &=\frac{\sin ^{2}\alpha \cos ^{4}\gamma }{%
2\sqrt{2}\pi }(\cos \alpha \sin \beta +\cos \beta \tan \gamma )(\cos \alpha
\cos \beta -\sin \beta \tan \gamma ),  \notag \\
\lbrack \Sigma _{2,1}^{(1)}]_{48} &=-\frac{\sin \alpha }{4\sqrt{2}\pi }%
(\cos \alpha \sin \beta \cos \gamma +\cos 2\alpha \cos \beta \sin \gamma
)(\cos \alpha \sin \beta \sin 2\gamma -\cos \beta \cos 2\gamma ),  \notag \\
\lbrack \Sigma _{2,1}^{(1)}]_{88} &=-\frac{\sin ^{2}\alpha }{32\sqrt{2}\pi }%
\{[(3+\cos 2\alpha )\cos 4\gamma +2\sin ^{2}\alpha ]\sin 2\beta +4\cos
\alpha \cos 2\beta \sin 4\gamma \},  \notag \\
\lbrack \Sigma _{2,1}^{(1)}]_{99} &=-\frac{\sin ^{2}\alpha \sin ^{4}\gamma
}{2\sqrt{2}\pi }(\cos \beta \cot \gamma -\cos \alpha \sin \beta )(\cos
\alpha \cos \beta +\sin \beta \cot \gamma ).\nonumber
\end{align}%
The nonzero elements of $\Sigma_{2,1}^{(2)}$ are
\begin{align}
\lbrack \Sigma _{2,1}^{(2)}]_{41} &=-\frac{1}{32\pi }(8\sin \alpha \cos
^{2}\alpha \sin ^{2}\beta \cos \gamma +\sin 4\alpha \sin 2\beta \sin \gamma
),  \notag \\
\lbrack \Sigma _{2,1}^{(2)}]_{81} &=\frac{\sin ^{2}\alpha \cos \alpha \sin
\beta }{4\pi }(\sin \beta \cos 2\gamma +\cos \alpha \cos \beta \sin 2\gamma
),  \notag \\
\lbrack \Sigma _{2,1}^{(2)}]_{43} &=\frac{1}{32\sqrt{2}\pi }[2(\cos \alpha
+\cos 3\alpha )(\cos 2\beta \cos 2\gamma -1)-(2+\cos 2\alpha +\cos 4\alpha
)\sin 2\beta \sin 2\gamma ],  \notag \\
\lbrack \Sigma _{2,1}^{(2)}]_{83} &=\frac{\sin \alpha }{4\sqrt{2}\pi }(\cos
2\alpha \sin \beta \cos \gamma +\cos \alpha \cos \beta \sin \gamma )(\sin
\beta \cos 2\gamma +\cos \alpha \cos \beta \sin 2\gamma ),  \notag \\
\lbrack \Sigma _{2,1}^{(2)}]_{94} &=\frac{\sin \alpha \sin \gamma }{4\pi }%
(\sin \beta \cos \gamma +\cos \alpha \cos \beta \sin \gamma )(\cos 2\alpha
\sin \beta \sin \gamma -\cos \alpha \cos \beta \cos \gamma ),  \notag \\
\lbrack \Sigma _{2,1}^{(2)}]_{47} &=\frac{\sin \alpha \cos \gamma }{4\pi }%
(\cos \alpha \sin \beta \cos \gamma +\cos \beta \sin \gamma )(\cos \alpha
\sin \beta \cos \gamma +\cos 2\alpha \cos \beta \sin \gamma ),  \notag \\
\lbrack \Sigma _{2,1}^{(2)}]_{87} &=-\frac{\sin ^{2}\alpha \cos \gamma }{%
4\pi }(\cos \alpha \sin \beta \cos \gamma +\cos \beta \sin \gamma )(\sin
\beta \cos 2\gamma +\cos \alpha \cos \beta \sin 2\gamma ),  \notag \\
\lbrack \Sigma _{2,1}^{(2)}]_{98} &=\frac{\sin ^{2}\alpha \sin \gamma }{%
4\pi }(\sin \beta \cos \gamma +\cos \alpha \cos \beta \sin \gamma )(\cos
\beta \cos 2\gamma -\cos \alpha \sin \beta \sin 2\gamma ).\nonumber
\end{align}%
The nonzero elements of $\Sigma_{2,1}^{(3)}$ are
\begin{align}
\lbrack \Sigma _{2,1}^{(3)}]_{91} &=-\frac{\sin ^{2}\alpha \cos \alpha \sin
\beta \sin \gamma }{2\sqrt{2}\pi }(\sin \beta \cos \gamma +\cos \alpha \cos
\beta \sin \gamma ),  \notag \\
\lbrack \Sigma _{2,1}^{(3)}]_{93} &=-\frac{\sin \alpha \sin \gamma }{4\pi }%
(\sin \beta \cos \gamma +\cos \alpha \cos \beta \sin \gamma )(\cos 2\alpha
\sin \beta \cos \gamma +\cos \alpha \cos \beta \sin \gamma ),  \notag \\
\lbrack \Sigma _{2,1}^{(3)}]_{97} &=\frac{\sin ^{2}\alpha \sin 2\gamma }{4%
\sqrt{2}\pi }(\cos \alpha \sin \beta \cos \gamma +\cos \beta \sin \gamma
)(\sin \beta \cos \gamma +\cos \alpha \cos \beta \sin \gamma ).\nonumber
\end{align}%
The nonzero elements of $\Sigma_{2,2}^{(-1)}$ are
\begin{align}
\lbrack \Sigma _{2,2}^{(-1)}]_{16} &=-\frac{1}{2\sqrt{2}\pi }\sin
^{2}\alpha \cos \alpha \cos \beta \sin \gamma ,  \notag \\
\lbrack \Sigma _{2,2}^{(-1)}]_{36} &=\frac{\sin \alpha \sin \gamma }{4\pi }%
(\cos \alpha \sin \beta \sin \gamma -\cos 2\alpha \cos \beta \cos \gamma ),
\notag \\
\lbrack \Sigma _{2,2}^{(-1)}]_{76} &=\frac{\sin ^{2}\alpha \sin 2\gamma }{4%
\sqrt{2}\pi }(\cos \alpha \cos \beta \cos \gamma -\sin \beta \sin \gamma ).\nonumber
\end{align}%
The nonzero elements of $\Sigma_{2,2}^{(0)}$ are
\begin{align}
\lbrack \Sigma _{2,2}^{(0)}]_{12} &=-\frac{\sin \alpha \cos ^{2}\alpha \cos
\beta }{2\sqrt{2}\pi },  \notag \\
\lbrack \Sigma _{2,2}^{(0)}]_{32} &=\frac{1}{8\pi }[2\cos ^{2}\alpha \sin
\beta \sin \gamma -(\cos \alpha +\cos 3\alpha )\cos \beta \cos \gamma ],
\notag \\
\lbrack \Sigma _{2,2}^{(0)}]_{72} &=\frac{\sin 2\alpha \cos \gamma }{4\sqrt{%
2}\pi }(\cos \alpha \cos \beta \cos \gamma -\sin \beta \sin \gamma ),  \notag
\\
\lbrack \Sigma _{2,2}^{(0)}]_{15} &=\frac{\sin ^{2}\alpha \cos \alpha \cos
\beta \cos \gamma }{2\sqrt{2}\pi },  \notag \\
\lbrack \Sigma _{2,2}^{(0)}]_{35} &=\frac{\sin \alpha \cos \gamma }{4\pi }%
(\cos 2\alpha \cos \beta \cos \gamma -\cos \alpha \sin \beta \sin \gamma ),
\notag \\
\lbrack \Sigma _{2,2}^{(0)}]_{75} &=\frac{\sin ^{2}\alpha \cos ^{3}\gamma }{%
2\sqrt{2}\pi }(\sin \beta \tan \gamma -\cos \alpha \cos \beta ),  \notag \\
\lbrack \Sigma _{2,2}^{(0)}]_{46} &=\frac{\sin \alpha \sin \gamma }{4\pi }%
(\cos \alpha \sin \beta \cos \gamma +\cos 2\alpha \cos \beta \sin \gamma ),
\notag \\
\lbrack \Sigma _{2,2}^{(0)}]_{86} &=-\frac{\sin ^{2}\alpha \sin \gamma }{%
4\pi }(\sin \beta \cos 2\gamma +\cos \alpha \cos \beta \sin 2\gamma ).\nonumber
\end{align}%
The nonzero elements of $\Sigma_{2,2}^{(1)}$ are
\begin{align}
\lbrack \Sigma _{2,2}^{(1)}]_{42} &=\frac{2\cos ^{2}\alpha \sin \beta \cos
\gamma +(\cos \alpha +\cos 3\alpha )\cos \beta \sin\gamma}{8\pi },
\notag \\
\lbrack \Sigma _{2,2}^{(1)}]_{82} &=-\frac{\sin 2\alpha }{8\pi }(\sin \beta
\cos 2\gamma +\cos \alpha \cos \beta \sin 2\gamma ),  \notag \\
\lbrack \Sigma _{2,2}^{(1)}]_{45} &=-\frac{\sin \alpha \cos \gamma }{4\pi }%
(\cos \alpha \sin \beta \cos \gamma +\cos 2\alpha \cos \beta \sin \gamma ),
\notag \\
\lbrack \Sigma _{2,2}^{(1)}]_{85} &=\frac{\sin ^{2}\alpha \cos \gamma }{%
4\pi }(\sin \beta \cos 2\gamma +\cos \alpha \cos \beta \sin 2\gamma ),
\notag \\
\lbrack \Sigma _{2,2}^{(1)}]_{96} &=\frac{\sin ^{2}\alpha \sin ^{3}\gamma }{%
2\sqrt{2}\pi }(\cos \alpha \cos \beta +\sin \beta \cot \gamma ).\nonumber
\end{align}%
The nonzero elements of $\Sigma_{2,2}^{(2)}$ are
\begin{align}
\lbrack \Sigma _{2,2}^{(2)}]_{92} &=\frac{\sin 2\alpha \sin \gamma }{4\sqrt{%
2}\pi }(\sin \beta \cos \gamma +\cos \alpha \cos \beta \sin \gamma ),  \notag
\\
\lbrack \Sigma _{2,2}^{(2)}]_{95} &=-\frac{\sin ^{2}\alpha \sin 2\gamma }{4%
\sqrt{2}\pi }(\sin \beta \cos \gamma +\cos \alpha \cos \beta \sin \gamma ).\nonumber
\end{align}

\section{Coefficients in the effective potential Eq.~\eqref{Veff}}\label{appeffw}
Here we provide the explicit expressions for the coefficients $c_{00}$, $c_{11}$, and $c_{22}$ in terms of the three Euler angles and the single molecule energy levels $\{E_{+},E_{-1},E_{-},E_{0}\}$. Here, the three coefficients can be derived from the perturbation theory as
\begin{align}
w_{0} &=-\frac{1}{288\pi ^{2}}\left\{\frac{\sin ^{2}\alpha \lbrack 3\cos
^{2}\alpha \sin 2\beta \cos \gamma +\frac{1}{2}(\cos \alpha +\cos 3\alpha
)(3\cos 2\beta -1)\sin \gamma ]^{2}}{E_{0}-E_{+}}\right.  \notag\\
&\quad\qquad\qquad+\frac{\cos ^{2}\alpha \sin ^{2}\alpha \lbrack \cos 2\alpha (1-3\cos
2\beta )\cos \gamma +3\cos \alpha \sin 2\beta \sin \gamma ]^{2}}{E_{-}-E_{+}}  \notag \\
&\quad\qquad\qquad+\frac{\sin ^{4}\alpha \cos ^{2}\alpha \sin ^{2}\gamma \lbrack 3\sin
2\beta \cos \gamma +\cos \alpha (3\cos 2\beta -1)\sin \gamma ]^{2}}{%
E_{0}-E_{+}}  \notag \\
&\quad\qquad\qquad+\frac{\sin ^{4}\alpha \cos ^{2}\alpha \lbrack 3\sin 2\beta \cos 2\gamma
+\cos \alpha (3\cos 2\beta -1)\sin 2\gamma ]^{2}}{E_{0}+E_{-}-2E_{+}}  \notag \\
&\quad\qquad\qquad\left.+\frac{\sin ^{4}\alpha \cos ^{4}\alpha \cos ^{4}\gamma (1-3\cos 2\beta
+3\sec \alpha \sin 2\beta \tan \gamma )^{2}}{E_{-}-E_{+}}\right\},
\end{align}

\begin{align}
w_{1} &=\frac{\sin ^{2}\alpha \cos ^{4}\alpha \sin ^{2}\beta }{4\pi
^{2}(E_{+}-E_{-1}-\omega )}+\frac{\sin ^{4}\alpha \cos
^{2}\alpha \sin ^{2}\beta \cos ^{2}\gamma }{4\pi ^{2}(2E_{+}-E_{-}-E_{-1}-\omega )}  \notag \\
&\quad+\frac{\sin ^{4}\alpha \cos ^{2}\alpha \sin ^{2}\beta \sin ^{2}\gamma }{4\pi ^{2}(2E_{+}-E_{0}-E_{-1}-\omega )}+\frac{\sin ^{4}\alpha \cos ^{2}\alpha \cos ^{2}\beta \cos ^{2}\gamma (\cos \alpha
\sin \beta \cos \gamma +\cos \beta \sin \gamma )^{2}}{2\pi ^{2}(2E_{+}-2E_{-}+\omega )}  \notag \\
&\quad+\frac{\sin ^{4}\alpha \cos ^{2}\alpha \sin ^{2}\beta \sin ^{2}\gamma
(\sin \beta \cos \gamma +\cos \alpha \cos \beta \sin \gamma )^{2}}{2\pi^{2}(2E_{+}-2E_{0}-\omega )}+\frac{\sin ^{4}\alpha \cos^{2}\alpha \sin ^{2}\beta \cos ^{2}\gamma (\cos \alpha \cos \beta \cos
\gamma -\sin \beta \sin \gamma )^{2}}{2\pi^{2}(2E_{+}-2E_{-}-\omega )}  \notag \\
&\quad+\frac{\sin ^{4}\alpha \cos ^{2}\alpha \cos ^{2}\beta \sin ^{2}\gamma
(\cos \beta \cos \gamma -\cos \alpha \sin \beta \sin \gamma )^{2}}{2\pi^{2}(2E_{+}-2E_{0}+\omega )}+\frac{\sin ^{2}\alpha \cos^{2}\alpha \cos ^{2}\beta (\cos 2\alpha \sin \beta \cos \gamma +\cos \alpha
\cos \beta \sin \gamma )^{2}}{4\pi ^{2}(E_{+}-E_{-}+\omega )}  \notag \\
&\quad+\frac{\sin ^{2}\alpha \cos ^{2}\alpha \sin ^{2}\beta (\cos 2\alpha \cos
\beta \sin \gamma +\cos \alpha \sin \beta \cos \gamma )^{2}}{4\pi^{2}(E_{+}-E_{0}-\omega )}+\frac{\sin ^{2}\alpha \cos
^{2}\alpha \sin ^{2}\beta \left( \cos 2\alpha \cos \beta \cos \gamma -\cos
\alpha \sin \beta \sin \gamma \right) ^{2}}{4\pi^{2}(E_{+}-E_{-}-\omega )}  \notag \\
&\quad+\frac{\sin ^{4}\alpha \cos ^{2}\alpha \sin ^{2}\beta \left( \sin \beta
\cos 2\gamma +\cos \alpha \cos \beta \sin 2\gamma \right) ^{2}}{4\pi
^{2}(2E_{+}-E_{0}-E_{-}-\omega )}+\frac{\sin
^{2}\alpha \cos ^{2}\alpha \cos ^{2}\beta \left( \cos 2\alpha \sin \beta
\sin \gamma -\cos \alpha \cos \beta \cos \gamma \right) ^{2}}{4\pi
^{2}(E_{+}-E_{0}+\omega )}  \notag \\
&\quad+\frac{\sin ^{4}\alpha \cos ^{2}\alpha \cos ^{2}\beta \left( \cos \alpha
\sin \beta \sin 2\gamma -\cos 2\gamma \cos \beta \right) ^{2}}{4\pi
^{2}(2E_{+}-E_{0}-E_{-}+\omega )},
\end{align}
and
\begin{align}
w_{2}=\frac{\cos ^{4}\alpha \cos ^{2}\beta \sin ^{2}\alpha }{8\pi
^{2}(E_{+}-E_{-1})}+\frac{\sin ^{4}\alpha \cos ^{2}\alpha
\cos ^{2}\beta \cos ^{2}\gamma }{8\pi ^{2}(2E_{+}-E_{-}-E_{-1})}+\frac{\sin ^{4}\alpha \cos ^{2}\alpha \cos ^{2}\beta
\sin ^{2}\gamma }{8\pi ^{2}(2E_{+}-E_{0}-E_{-1})}.
\end{align}

\section{Induced gauge potentials}\label{appgauge}

In the main text, the effective potential is derived as an approximation of the adiabatic potential connecting to the $|++\rangle$ state. Although the validity of this effective potential is justified by comparing the two-body scattering via effective potential with the multichannel calculations, here we further validate the effective potential by calculating the effective potential with induced gauge potentials in the Floquet space. Following the standard procedure, the BO approximation Hamiltonian for the relative motion of two $|+\rangle$ molecules in the $n=0$ Floquet sector takes the form
\begin{equation}
H_{\mathrm{BO}}=\frac{1}{M}\left[{\boldsymbol p}-{\boldsymbol A}({\boldsymbol r})\right]^{2}+V_{\mathrm{sc}}({\boldsymbol r})+V_{\mathrm{0,1}}^{(\mathrm{ad})}({\boldsymbol r}),
\end{equation}
where 
\begin{align}
{\boldsymbol A}({\boldsymbol r})=i\langle V_{0,1}^{(\mathrm{ad})}({\boldsymbol r})|\nabla
|V_{0,1}^{(\mathrm{ad})}({\boldsymbol r})\rangle,
\end{align}
is the vector potential and
\begin{align}
V_{\mathrm{sc}}({\boldsymbol r})=\frac{1}{M}\left.\sum_{n,\nu}\right.^{\prime}\left| \left\langle V_{n,\nu }^{(\mathrm{ad})}({\boldsymbol r})\left| \nabla
\right| V_{0,1}^{(\mathrm{ad})}({\boldsymbol r})\right\rangle \right|^{2},
\end{align}
is the scalar potential with the term of $(n,\nu)=(0,1)$ being excluded in the primed sum.

To compute the vector and scalar potentials, let us get a closer look at the adiabatic eigenstates. As shown previously, due to the conservation of the projection of the total angular momentum,  $|V_{0,1}^{(\mathrm{ad})}({\boldsymbol r})\rangle$ possesses a definite projection quantum number, say $m'$. We can then expand it as
\begin{align}
|V_{0,1}^{(\mathrm{ad})}({\boldsymbol r})\rangle &=\sum_{{n,}\nu }e^{i({m_{\nu}+n})\varphi }f_{{n,}\nu }(r,\theta )\left\vert {n,}\nu \right\rangle\nonumber\\
&=e^{im'\varphi}\sum_{{n,}\nu }e^{i({m_{\nu}^{(0)}+n})\varphi }f_{{n,}\nu }(r,\theta )\left\vert {n,}\nu \right\rangle
\label{V01}
\end{align}
where, for $\nu=1$ to $9$, $m_\nu^{(0)}=0,2,0,1,2,3,0,1$, and $2$, respectively, representing the simplest set of $\{m_\nu\}$. Moreover, $f_{{n,}\nu }(r,\theta )$ are real functions and satisfy the normalization condition $\sum_{n,\nu}f_{n,\nu}^2(r,\theta)=1$ for any $r$ and $\theta$. This representation (\ref{V01}) makes it particularly convenient to work within the spherical coordinate. A straightforward calculation immediately leads to the vector potential $${\boldsymbol A}({\boldsymbol r})=A_{\varphi }(r,\theta )\hat{\boldsymbol \varphi},$$ where
\begin{equation}
A_{\varphi }(r,\theta )=-\frac{1}{r\sin \theta }\sum_{n,\nu }({m_{\nu}^{(0)}+n})f_{{n,}\nu }^{2}(r,\theta)-\frac{m'}{r\sin\theta}.  \label{A}
\end{equation}%
Subsequently, the Berry curvature ${\boldsymbol B}({\boldsymbol r})=\nabla \times {\boldsymbol A}({\boldsymbol r})$ becomes
\begin{align}
{\boldsymbol B}({\boldsymbol r})&=B_{r}(r,\theta)\hat{\boldsymbol r}+B_{\theta }(r,\theta )\hat{\boldsymbol \theta},
\end{align}
where $B_r=(r\sin\theta)^{-1}\partial_\theta(A_\varphi\sin\theta)$ and $B_\theta=-r^{-1}\partial_r(rA_\varphi)$. It can be readily verified that the last term in Eq.~\eqref{A} does not contribute to the Berry curvature, suggesting that the the $m'$ term in Eq.~\eqref{A} only represents a pure gauge. Therefore, without loss of generality, we can set $m'=0$ and choose the simplest set $\{m_\nu^{(0)}\}$ for the projection quantum numbers.

To proceed further, we introduce, for $(n,\nu)\neq (0,1)$ and $\alpha =r,\theta ,\varphi $, the quantities
\begin{align}
D_{n,\nu }^{\alpha } &\equiv\left\langle V_{n,\nu}^{(\mathrm{ad})}({\boldsymbol r})\left| \partial_{\alpha }\right| V_{0,1}^{(\mathrm{ad})}({\boldsymbol r})\right\rangle=\frac{\left\langle V_{n,\nu }^{(\mathrm{ad})}({\boldsymbol r})\left\vert \partial _{\alpha}{\mathbf V}({\boldsymbol r})\right\vert V_{0,1}^{(\mathrm{ad})}({\boldsymbol r})\right\rangle }{V_{0,1}^{(\mathrm{ad})}({\boldsymbol r})-V_{n,\nu }^{(\mathrm{ad})}({\boldsymbol r})},
\end{align}
which, in analog to derivation of the Kubo formula, can be obtained by differentiating the eigenequation~\eqref{V}. The components of the induced magnetic field can then be expressed as
\begin{eqnarray}
B_{r}(r,\theta ) &=&i\frac{1}{r^{2}\sin \theta }\left.\sum_{n,\nu}\right.^{\prime}\varepsilon _{r\alpha \beta }D_{n,\nu }^{\alpha \ast }D_{n,\nu
}^{\beta },  \label{Br}\\
B_{\theta }(r,\theta ) &=&i\frac{1}{r\sin \theta }\left.\sum_{n,\nu}\right.^{\prime}\varepsilon _{\theta \alpha \beta }D_{n,\nu }^{\alpha \ast }D_{n,\nu
}^{\beta },  \label{Bt}
\end{eqnarray}
where $\varepsilon_{\alpha\beta\gamma}$ is the Levi-Civita symbol with $\alpha,\beta,\gamma$ being a permutation of $r,\theta,\varphi$. Moreover, the scalar potential can be alternatively expressed as
\begin{align}
V_{\mathrm{sc}}({\boldsymbol r})&=\frac{1}{M}\left.\sum_{n,\nu}\right.^{\prime}\left\vert \frac{\left\langle V_{n,\nu }^{(\mathrm{ad})}({\boldsymbol r})\left\vert
\nabla {\mathbf V}({\boldsymbol r})\right\vert V_{0,1}^{(\mathrm{ad})}({\boldsymbol r})\right\rangle }{V_{0,1}^{(\mathrm{ad})}({\boldsymbol r})-V_{n,\nu }^{(\mathrm{ad})}({\boldsymbol r})}\right\vert ^{2}.  \label{Vsc}
\end{align}%
Using Eqs.~(\ref{Br})-(\ref{Vsc}), the magnetic field and the scalar potential can be more efficiently calculated since we may analytically compute the derivatives of the interaction potential.

We can now estimate the contributions due to the induced gauge fields. To this end, we first note that, based on the first-order perturbation theory, the components $f_{{n,}\nu }(r,\theta )$
scale as $1/r^{3}$ for $(n,\nu)\neq (0,1)$. Consequently, $A_{\varphi
}({\boldsymbol r})$ decays as $1/r^{7}$, which leads to the magnetic field ${\boldsymbol B}({\boldsymbol r})\sim 1/r^{8}$. The scaling behavior of ${\boldsymbol B}({\boldsymbol r})$ can also be derived directly from Eqs.~\eqref{Br} and \eqref{Bt}, where $r\partial _{r}{\mathbf V}({\boldsymbol r})$ and $\partial_{\theta,\varphi }{\mathbf V}({\boldsymbol r})$ scale as $1/r^{3}$. Similarly, the scalar potential $V_{\mathrm{sc}}({\boldsymbol r})$ scales as $1/r^{8}$, as indicated by Eq. (\ref{Vsc}). Recall the $r$ dependence of the effective potential $V_{\rm eff}$, these estimations clearly indicate that the induced gauge fields are of importance only at the short distance.

To quantitatively measure the effects of the gauge potentials, we compare the scatterings governed by $H_{\rm BO}$ with those governed by $H_{\rm eff}$. For this purpose, we introduce the total potential
\begin{align}
V_{\rm tot}({\boldsymbol r})&=-\frac{2{\boldsymbol A}\cdot {\boldsymbol p}-{\boldsymbol A}^2}{M}+V_{\rm sc}({\boldsymbol r})+V_{\rm eff}({\boldsymbol r}),\label{vtot}
\end{align}
such that the BO Hamiltonian can now be rewritten as 
\begin{align}
H_{\rm BO}\approx\frac{{\boldsymbol p}^2}{M}+V_{\rm tot}({\boldsymbol r}). 
\end{align}
In Eq.~\eqref{vtot}, we have replaced $V_{0,1}^\mathrm{(ad)}({\boldsymbol r})$ by $V_{\rm eff}({\boldsymbol r})$, which, as shown in Fig.~\ref{potcalib}, is a very good approximation if $C_3$ and $C_6$ are obtained by fitting. Again, we solve the equation $H_{\rm BO}\psi({\boldsymbol r})=(k_{01}^2/M)\psi({\boldsymbol r})$ by utilizing the expansion $\psi ({\boldsymbol r})=\sum_{lm}r^{-1}\phi_{lm}(r)Y_{lm}(\hat{\boldsymbol r})$. Here the radial wave functions satisfy
\begin{align}
-\frac{1}{M}&\left[\frac{d^2}{dr^2}-\frac{l(l+1)}{r^2}\right]\phi_{lm}(r)+\sum_{l'm'}[V_{\rm tot}^{(m)}(r)]_{ll'}\delta_{mm'}\phi _{l'm'}(r)=\frac{k_{01}^2}{M} \phi_{lm}(r),
\label{SEl}
\end{align}
where the interaction matrix element 
\begin{align}
[V^{(m)}_{\rm tot}(r)]_{ll'} &=\int d\hat{\boldsymbol r}Y_{lm}^{\ast}(\hat{\boldsymbol r})V_{\rm tot}({\boldsymbol r})Y_{l^{\prime }m}(\hat{\boldsymbol r})\nonumber\\
&=\int d\hat{\boldsymbol r}Y_{lm}^{\ast}(\hat{\boldsymbol r})V^{(m)}_{\rm tot}(r,\theta)Y_{l^{\prime }m}(\hat{\boldsymbol r}) \label{Vm} 
\end{align}
with
\begin{align}
V_{\rm tot}^{(m)}(r,\theta)&=-\frac{2mA_{\varphi }(r,\theta )}{Mr\sin \theta 
}+\frac{A_\varphi^{2}(r,\theta)}{M}+V_{\mathrm{sc}}(r,\theta)+V_{\rm eff}(r,\theta)\label{vtotm}
\end{align}%
being a potential that depends on the magnetic quantum number $m$, in contrast to Eq.~\eqref{mateleveff}. 

\begin{figure}[tbp]
\includegraphics[width=0.4\linewidth]{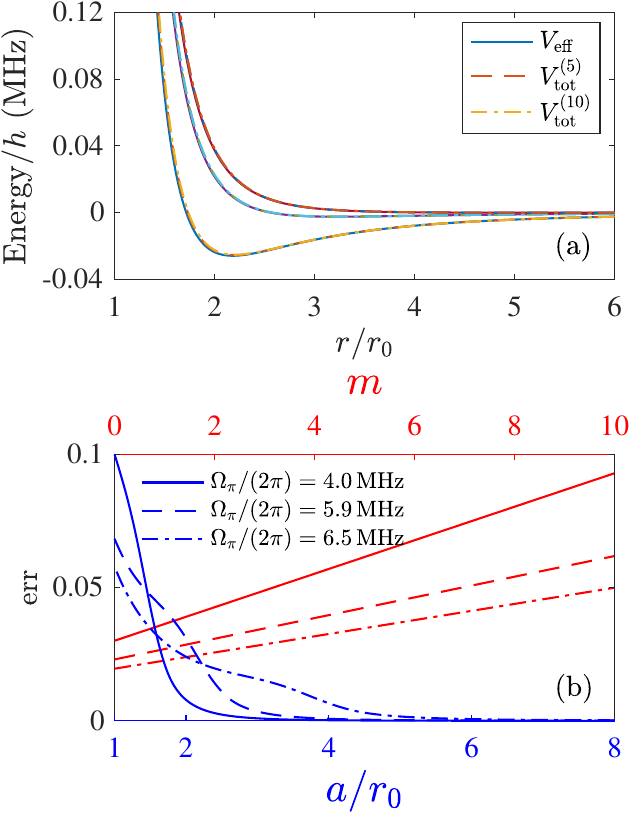}
\caption{(a) Comparison of the adiabatic potential $V_{\rm eff}(r,\pi/2)$ with the total potentials $V_{\rm tot}^{(m)}(r,\pi/2)$ with $m=5$ and $10$ for $\Omega_\pi/(2\pi)=6.5$, $5.9$, and $4\,{\rm MHz}$ (for three sets of curves in descending order). (b) Relative error versus $m$ and $a$ for different $\Omega_\pi$'s.}
\label{gauerr}
\end{figure}

In Fig.~\ref{gauerr}(a), we compare the effect potential $V_{\rm eff}$ with the total potential $V_{\rm tot}^{(m)}$ with $m=5$ and $10$ for various $\Omega_\pi$'s. As can be seen, the discrepancy at large distance is zero and can only be seen inside the shield core. Although the discrepancy increases as $m$ grows, the gauge potential correction remains minor even for $m=10$. To quantify the discrepancy between $V_{\rm tot}^{(m)}$ and $V_{\rm eff}$, we define the relative error of the potentials as
\begin{align}
{\rm err}(m)\equiv\frac{\int_{a}^{b}\left|V_{\rm tot}^{(m)}(r,\theta)-V_{\rm eff}(r,\theta)\right|dr}{\int_{a}^{b}\left|V_{\rm eff}(r,\theta)\right|dr},
\end{align}
where the upper integration limit is fixed at $b=10r_0$ and the lower limit $a$ is changeable. Figure~\ref{gauerr}(b) plot the relative error as a function of $m$ for different $\Omega_\pi$'s. Here $a$ is so chosen that $V_{\rm eff}(a,\pi/2)/h=1\,{\rm MHz}$, an energy scale that is sufficiently large. As can be seen, $\mathrm{err}(m)$ monotonically grows with $m$. In particular, for $\Omega_\pi/(2\pi)=4\,{\rm MHz}$, the relative error is around $10\%$ for $m=10$, which seems to suggest that the adiabatic potential becomes invalid for large $m$. To identify the origin of the error, we also plot, in Fig.~\ref{gauerr}(b), the relative error as a function $a$ for $m=10$, showing that $\mathrm{err}$ quickly drops as $a$ increases. This observation clearly indicates that the relative error is mainly contributed by discrepancy at short distance, which, as shall be shown shortly, can be screened out by the centrifugal barrier at low temperature.

Finally, we point out that a careful examination of the inset of Fig.~\ref{potcalib}(b) further reveals a tiny discrepancy between $a_{01l,01l}^{(m)}$ with $\pm m$, in contrast to $a_{l,l}^{(m)}$ that is independent of the sign of $m$. This tiny discrepancy can be easily understood by the first term of the gauge potential [see Eq.~\eqref{vtotm}] which explicitly breaking time-reversal symmetry which, in contrast, is preserved in $H_{\mathrm{eff}}$.

\end{document}